\documentclass[12pt]{article}
\usepackage{jheppub}
\usepackage{amsmath,amssymb,amsfonts,graphicx,subfigure,
slashed,amsthm,mathtools,upgreek,enumerate,tensor}
\usepackage[dvipsnames]{xcolor}
\usepackage{arydshln}
\usepackage{braket}
\usepackage{color}
\usepackage{cases}

\usepackage{hyperref}
\usepackage[utf8]{inputenc}
\usepackage[titletoc]{appendix}

\definecolor{phthaloblue}{rgb}{0.0, 0.06, 0.54}
\definecolor{purple}{rgb}{0.5 ,0, 0.7}
\definecolor{bluegreen}{rgb}{0, 0.45, 0.35}
\definecolor{sakura}{rgb}{1 ,0.52, 0.74}
\definecolor{wakakusa}{rgb}{0.45 ,0.74, 0}
\definecolor{brown}{rgb}{0.48 ,0.23, 0}
\definecolor{skyblue}{rgb}{0.21 ,0.7, 1.}
\definecolor{purplegray}{rgb}{0.35,0.35,0.73}
\usepackage{hyperref}
\hypersetup{colorlinks=true, linkcolor=bluegreen, citecolor=purplegray, urlcolor=purplegray}

\title{
No Smooth Spacetime in Lorentzian Quantum Cosmology and Trans-Planckian Physics
}

\author{\large Hiroki Matsui${}^a$, Shinji Mukohyama${}^{a,b}$, 
Atsushi Naruko${}^a$}

\emailAdd{hiroki.matsui@yukawa.kyoto-u.ac.jp} 
\emailAdd{shinji.mukohyama@yukawa.kyoto-u.ac.jp}
\emailAdd{naruko@yukawa.kyoto-u.ac.jp}
\affiliation{${}^a$Center for Gravitational Physics and Quantum Information,
Yukawa Institute for Theoretical Physics, Kyoto University,
Kitashirakawa Oiwakecho, Sakyo-ku,
Kyoto 606-8502, JAPAN\medskip\\
${}^b$Kavli Institute for the Physics and Mathematics of the Universe (WPI), The University of Tokyo Institutes for Advanced Study, The University of Tokyo, Kashiwa, Chiba 277-8583, Japan\medskip\\}

\abstract{
In minisuperspace quantum cosmology, the Lorentzian path integral formulations of the no-boundary and tunneling proposals have recently been analyzed.
But it has been pointed out that the wave function of linearized perturbations around a homogeneous and isotropic background is of an inverse Gaussian form and thus that their correlation functions are divergent. In this paper, we revisit this issue and consider the problem of perturbations in Lorentzian quantum cosmology by modifying the dispersion relation based on trans-Planckian physics. We consider two modified dispersion relations, the generalized Corley-Jacobson dispersion relation with higher momentum terms and the Unruh dispersion relation with a trans-Planckian mode cut-off, as examples.
We show that the inverse Gaussian problem of perturbations in Lorentzian quantum cosmology is hard to overcome with the trans-Planckian physics modifying the dispersion relation at short distances.}

\preprint{YITP-22-133, IPMU22-0058}

\begin{document}

\maketitle

\section{Introduction} 
Classical general relativity (GR) does not answer how the Planck-sized primordial universe was created and developed since quantum gravity will be necessary to address such questions. Quantum cosmology tries to describe how the universe was created based on the quantum gravity approach and introduces the so-called wave function of the universe, which is the wave-functional $\Psi[g]$ of the spatial metric $g$ induced on a $3$-geometry. In canonical quantum gravity, the wave function is given by a solution for the Wheeler-DeWitt equation with suitable boundary conditions. Alternatively, the wave function of the universe can be formulated by the path integral of gravity, $\Psi[g]=\int\mathcal{D} g^{(4)}\, e^{i S[g^{(4)}] / \hbar}$, where the $4$-dimensional metric $g^{(4)}$ is restricted to those inducing $g$ on the $3$-geometry and the diffeomorphism invariance is properly treated.

The most well-known formulations for the wave function of the universe $\Psi[g]$ are the no-boundary proposal~\cite{Hartle:1983ai} and the tunneling proposal~\cite{Vilenkin:1984wp}. 
The Lorentzian path integral formulation of these proposals has recently been analyzed 
in minisuperspace quantum cosmology~\cite{Feldbrugge:2017kzv,DiazDorronsoro:2017hti} 
and, there are some doubts about them 
under the inclusion of perturbations around a homogeneous and isotropic background~\cite{Feldbrugge:2017fcc,Feldbrugge:2017mbc}. 
The linearized perturbations around the background are described by a wave function that takes the form of inverse Gaussian, and thus the correlation functions of perturbations diverge. As a consequence, they will be out of control. 
This suggests that the anisotropy and inhomogeneity of spacetime caused by the linearized perturbations are not suppressed, and 
neither the no-boundary proposal nor the tunneling proposal is likely to be consistent with cosmological observations.
Also, we note that DeWitt's proposal~\cite{DeWitt:1967yk} which
states the vanishing wave function of the universe at the big-bang singularity suffers from the perturbation problem in GR~\cite{Matsui:2021yte,Martens:2022dtd}.

There are several attempts or approaches to address the linearized perturbation problems for the no-boundary proposal and the tunneling proposal in the literature~\cite{DiazDorronsoro:2018wro,Feldbrugge:2018gin,Vilenkin:2018dch,Vilenkin:2018oja,Wang:2019spw,Bojowald:2018gdt,DiTucci:2019dji,DiTucci:2019bui,Halliwell:2018ejl,Bojowald:2020kob, Lehners:2021jmv}. Motivated by the early success of the no-boundary proposal~\cite{Hartle:1983ai} based on the Euclidean path integrals, the authors in Refs.~\cite{DiazDorronsoro:2017hti,DiazDorronsoro:2018wro} proposed the integral of the lapse function in the complex plane and also a particular initial condition for the momentum conjugate to the scale factor. However, it has been claimed by Refs.~\cite{Feldbrugge:2017mbc,Feldbrugge:2018gin} that the linearized perturbations still have the inverse Gaussian wave function and thus divergent correlation functions. Recently, Refs.~\cite{DiTucci:2019dji,DiTucci:2019bui} proposed different boundary conditions for the no-boundary proposal to avoid the inverse Gaussian wave function for linearized perturbations at the price of abandoning the notion of a sum over compact and regular geometries. On the other hand, for the tunneling proposal, the authors in Refs.~\cite{Vilenkin:2018dch,Vilenkin:2018oja} introduced a specific boundary term for the gravitational action of the linearized perturbations to satisfy the Robin boundary condition. However, it is still unclear whether this proposal works in the presence of tensor perturbations: the boundary term for tensor perturbation should be consistently derived from a nonlinear boundary term that is defined in a geometrical way and that should apply to the background as well. Although loop quantum gravity might provide some insights on the problem and one might hope that dynamical signature change might ameliorate the behavior of the wave function of the background and perturbations in the UV regime, the result of Ref.~\cite{Bojowald:2018gdt,Bojowald:2020kob} does not ensure the IR stability of the perturbations.

In this {\it paper}, we first revisit the problem of the inverse Gaussian wave function in Lorentzian quantum cosmology in the context of GR. We shall clearly show that the inverse Gaussian problem for tensor perturbations is inevitable as far as one requires the regularity of the on-shell gravitational action. Upon using the equations of motion, it is shown that the on-shell action can be written as a sum of boundary contributions, one from the UV and the other from the IR. The UV contribution tends to diverge, and thus its regularity selects a particular mode function for the tensor perturbation up to an overall factor. This mode function inevitably leads to the inverse Gaussian wave function for the tensor perturbation. Then, we revisit the problem of linearized perturbations in Lorentzian quantum cosmology 
by assuming dispersion relations motivated 
by the so-called trans-Planckian physics~\cite{Martin:2000xs,Brandenberger:2000wr,Niemeyer:2000eh,Martin:2002kt,Ashoorioon:2004vm,Ashoorioon:2011eg}. The trans-Planckian physics should be expected to modify the dispersion relations for perturbations 
when the physical wavelength is smaller than the Planck scale. 
The modified dispersion relation was first introduced in the black hole physics~\cite{Unruh:1994je,Corley:1996ar} and then was applied to cosmology~\cite{Martin:2000xs,Brandenberger:2000wr,Niemeyer:2000eh}. The possibility that the dispersion relation is modified near the big-bang singularity is quite reasonable, and such effects are caused by quantum gravity. We consider the Unruh dispersion relation~\cite{Unruh:1994je} and generalized Corley-Jacobson dispersion relation~\cite{Corley:1996ar,Corley:1997pr} as examples of the modified dispersion relation and discuss whether such modified dispersion relation can solve the problem of the inverse Gaussian wave function for perturbations in Lorentzian quantum cosmology.

The rest of the present paper is organized as follows. 
In Section~\ref{sec:no-boundary-proposal}, 
we review the no-boundary and tunneling proposals 
based on the Lorentzian path integral formulation. 
In Section~\ref{sec:Full-analysis}, we revisit  
the problem of the inverse Gaussian wave function for linearized tensor perturbations in GR.
In Section~\ref{sec:TPP}, we consider the trans-Planckian physics and modified dispersion relations such as generalized Corley-Jacobson dispersion relation
and Unruh dispersion relation.
In Section~\ref{sec:conclusion} we conclude our work.

\section{No-boundary and tunneling propagator}
\label{sec:no-boundary-proposal}
In this section, we will briefly review the no-boundary and tunneling proposals based on the Lorentzian path integral. Boundary conditions for these proposals in quantum cosmology under the minisuperspace approximation can be implemented in the Lorentzian path integral~\cite{Feldbrugge:2017kzv,DiazDorronsoro:2017hti}, which is different from the Euclidean path integral formulation. Integrals of phase factors such as $e^{i S_{\rm GR}/\hbar}$ usually do not manifestly converge, but the convergence can be improved by shifting the contour of the integral onto the complex plane by applying Picard-Lefschetz  theory~\cite{Witten:2010cx}. According to Cauchy's theorem, if the integrand does not have poles in a region on the complex plane, the Lorentzian nature of the integral is preserved even if the integration contour on the complex plane is deformed within such a region. 
In particular, as we will show later, in minisuperspace quantum cosmology, the path integral can be rewritten as an integral depending only on the gauge-fixed lapse function $N$ under the semiclassical approximation.
The Lorentzian path integral in the no-boundary and tunneling proposals can be directly performed.

Let us consider a closed Friedmann-Lema\^{i}tre-Robertson-Walker (FLRW) universe with tensor-type metric perturbations whose line element is written as
\begin{equation}\label{eq:metric}
\mathrm{d} s^{2}= -N(t)^2 \mathrm{d}t^2 
+ a(t)^2\left[ \Omega_{ij} ({\bf x})+h_{i j} (t \,, {\bf x}) \right] \mathrm{d} x^{i} \mathrm{d} x^{j}\,,
\end{equation}
where $t$ is a time variable, $a(t)$ is the scale factor, $N(t)$ is the lapse function, $\Omega_{ij}$ is the metric of the unit 3-sphere, $h_{ij}$ represents the tensor perturbation satisfying the transverse and traceless condition, $\Omega ^{i j} h_{i j} = \Omega^{ki}D_k h_{i j} = 0$, $\Omega^{ij}$ is the inverse of $\Omega_{ij}$ and
 $D_i$ is the spatial covariant derivative compatible with $\Omega_{ij}$. 
Given this metric, the gravitational action is expanded up to the second order in the perturbation $h_{i j}$ as
 $S_{\rm GR} = S_{\rm GR}^{(0)} (h^0) +S_{\rm GR}^{(2)} (h^2) +\mathcal{O}(h^3)$:
\begin{align}
S_{\rm GR}^{(0)} &=  2\pi^2 \int N \mathrm{d}t \, \left[ -\frac{3}{N^2}{a(\dot{a})}^2 
+ 3a - \Lambda a^3 \right] \,, 
\label{action_0} \\
S_{\rm GR}^{(2)} &= 2\pi^2 \int Na \mathrm{d}t \, \sum_{snlm}\biggl[
\frac{a^2}{8N^2}\left(\dot{h}^{s}_{nlm}\right)^2
-\frac{1}{8}\left((n^2-3)+6 \right)(h^s_{nlm})^2\biggr]\,,
\label{action_2} 
\end{align}
where we take the Planck mass unit with $M_{\rm Pl}=1/\sqrt{8\pi G}=1$. We have expanded the tensor perturbation $h_{ij}$ in terms of the tensor hyper-spherical harmonics with each coefficient $h^s_{nlm}$ being a function of the time $t$, where $s=\pm$ is the polarization label, and the integers ($n$, $l$, $m$) run over the ranges $n\geq3$, $l \in [0,n-1]$, $m \in [-l,l]$~\cite{Gerlach:1978gy}. We will restrict our consideration to one mode of tensor perturbations and denote $h^s_{nlm}$ of our interest simply by $h$, suppressing the indices $snlm$.

Hereafter, we shall construct the gravitational propagator preserving reparametrization invariance through the Batalin-Fradkin-Vilkovisky (BFV) formalism~\cite{Fradkin:1975cq,Batalin:1977pb}.
For the gauge-fixing choice $\dot{N}=0$, 
the BFV path integral reads~\cite{Halliwell:1988wc},
\begin{align}
G[a,h]&\equiv \int \mathcal{D}a\mathcal{D}h\mathcal{D}p_a\mathcal{D}p_h\mathcal{D}\Pi\,\mathcal{D}N\,\mathcal{D}\rho\,\mathcal{D}\bar{c}\,\mathcal{D}\bar\rho\,\mathcal{D}c\,\exp(iS_{\rm BRS}/\hbar) ,\\
S_{\rm BRS}&\equiv \int_{t_0}^{t_1} {\rm d}t\left(p_{a}\dot{a}+p_{h}\dot{h}
- N \mathcal{H} + \Pi \dot{N} +\bar\rho\dot{c} +\bar{c}\dot\rho-\bar\rho\rho\right),
\end{align}
where $p_{a,h}$ are the momenta conjugate to $a,h$. Here, $S_{\rm BRS}$ is the Becchi-Rouet-Stora (BRS) invariant action, including the Hamiltonian constraint $\mathcal{H}(a,h)$, a Lagrange multiplier $\Pi$ and ghost fields $\rho$, $\bar{\rho}$, $c$, $\bar{c}$, preserving the BRS symmetry, i.e. invariant under the following transformation. 
\begin{align}
\begin{split}
\delta a=\lambda c\frac{\partial \mathcal{H} }{\partial p_{a}}\,,\;\;\delta p_{a}=-\lambda c\frac{\partial \mathcal{H} }{\partial a}\,,\;\;
\delta h=\lambda c\frac{\partial \mathcal{H} }{\partial p_{h}}\,,\;\;\delta p_{h}=-\lambda c\frac{\partial \mathcal{H} }{\partial h}\,,\\ 
\delta N= \lambda\rho\,,\;\;
\delta\bar{c}=-\lambda\Pi\,,\;\;\delta\bar\rho=-\lambda \mathcal{H} \,,\;\; \delta \Pi=\delta c=\delta \rho=0,
\end{split}
\end{align}
where $\lambda$ is a parameter.
The ghost and multiplier parts can be integrated out, and eventually, we obtain
the following gravitational propagator,
\begin{align}\label{G-propagator}
G[a,h] &= \int \! \mathrm{d}N(t_1-t_0)
\int \mathcal{D}a\mathcal{D}h  \mathcal{D}p_a\mathcal{D}p_h\exp\left(i \int_{t_0}^{t_1} {\rm d}t\left(p_{a}\dot{a}+p_{h}\dot{h}  - N \mathcal{H}\right)/ \hbar\right)\nonumber
\\& = \int\! \mathrm{d}N(t_1-t_0) \int 
\mathcal{D}a\mathcal{D}h \exp\left(i S_{{\rm GR}}[a,h,N]
/ \hbar\right)\,,
\end{align}
which is the integral over the proper time $N(t_1-t_0)$ 
between the initial and final configurations. 
The gravitational propagator for the no-boundary and tunneling proposals can be given by Eq.~\eqref{G-propagator} with which the integration is performed from a 3-geometry of zero sizes, i.e. nothing, to a finite one~\cite{Halliwell:1988ik}. 
To simplify the analysis, we
introduce the new time coordinate $\tau$ that is related to $t$ by 
\begin{equation}\label{new-time}
 a(t) \mathrm{d}t=\mathrm{d}\tau \,,
\end{equation}
We shall take the initial and final times as $\tau_0=0$ and $\tau_1=1$, respectively, and use this notation for the later analysis.

The gravitational propagator at the zeroth-order in perturbation is given in terms of the zeroth-order action in the same order as
\begin{equation}\label{G-propagator_0}
 G^{(0)}[q_1] = \int_{0,-\infty}^\infty \mathrm{d}N  \int_{q(\tau=\tau_0)=0}^{q(\tau=\tau_1)=q_1}
\mathcal{D}q  \, e^{i S_{\rm GR}^{(0)}[q,N] / \hbar} \,,
 \end{equation}  
where 
\begin{equation}\label{def-q}
q(\tau)=a^2(\tau) \,,
 \end{equation}  
and $q(\tau=0)=0$. 
Such a change of variable 
 alters the path integral measure,
but would not change the dominant behavior of the propagator.
In the semi-classical analysis, we obtain the following expression~\cite{Feldbrugge:2017kzv}, 
\begin{equation}\label{Path-integral2}
G^{(0)}[q_1]  =  \sqrt{\frac{3\pi i}{2\hbar}}\int_{0,-\infty}^\infty  \frac{\mathrm{d} N}{N^{1/2}} 
\exp \left(\frac{i S_{\rm on-shell}^{(0)}[N]}{\hbar}\right)\,,
\end{equation}
where $S_{\rm on-shell}^{(0)}[N]$ is the on-shell action for the background,
\begin{equation}\label{eq:semi-action1}
S_{\rm on-shell}^{(0)}[N]=2\pi^2 \left[ 
\frac{N^3H^4}{4} + N \left( -\frac{3H^2 q_1}{2} +3 \right) +\frac{1}{N}\left( -\frac{3}{4}q_1^2\right) \right]\,.
\end{equation}
Note that $S_{\rm on-shell}^{(0)}[N]$ has four saddle points $N_s$, 
found by demanding ${\mathrm{d} S_{\rm on-shell}^{(0)}[N]\over \mathrm{d} N} =0$,
\begin{equation}\label{eq:saddle-points}
N_s=\frac{c_1}{H^2}
\left[c_2i+\left(q_{1}H^2-1\right)^{1/2}\right],
\end{equation}
with $c_{1,2} \in \{-1 , +1\}$ and $H^2\equiv \Lambda/3$.
Then the saddle-point action $S_{\rm on-shell}^{(0)}[N_s] $ reads,
\begin{align}\label{eq:saddle-point-action}
S_{\rm on-shell}^{(0)}[N_s] &=c_1\frac{4\pi^2}{H^2}  \left[c_1 i - \left(q_1H^2-1\right)^{3/2} \right].
 \end{align}
The integration over $N$ in the expression \eqref{Path-integral2} of the propagator can be performed by using the Picard-Lefschetz theory~\cite{Witten:2010cx} after finding the relevant saddle points and the steepest descent contours in the complex $N$-plane.
When we integrate the lapse function over $N \in (0, \infty)$, this propagator leads to the tunneling wave function. On the other hand, integrating the lapse function over $N \in (-\infty, \infty)$~\cite{DiazDorronsoro:2017hti} in \eqref{Path-integral2} provides either the real part of the tunneling wave function or the no-boundary wave function as found in Ref.~\cite{Hartle:1983ai} with the Euclidean path integral method, depending on whether the path goes above or below the singularity at $N = 0$.

\begin{figure}[t] 
\centering
\includegraphics[width=0.95\textwidth]{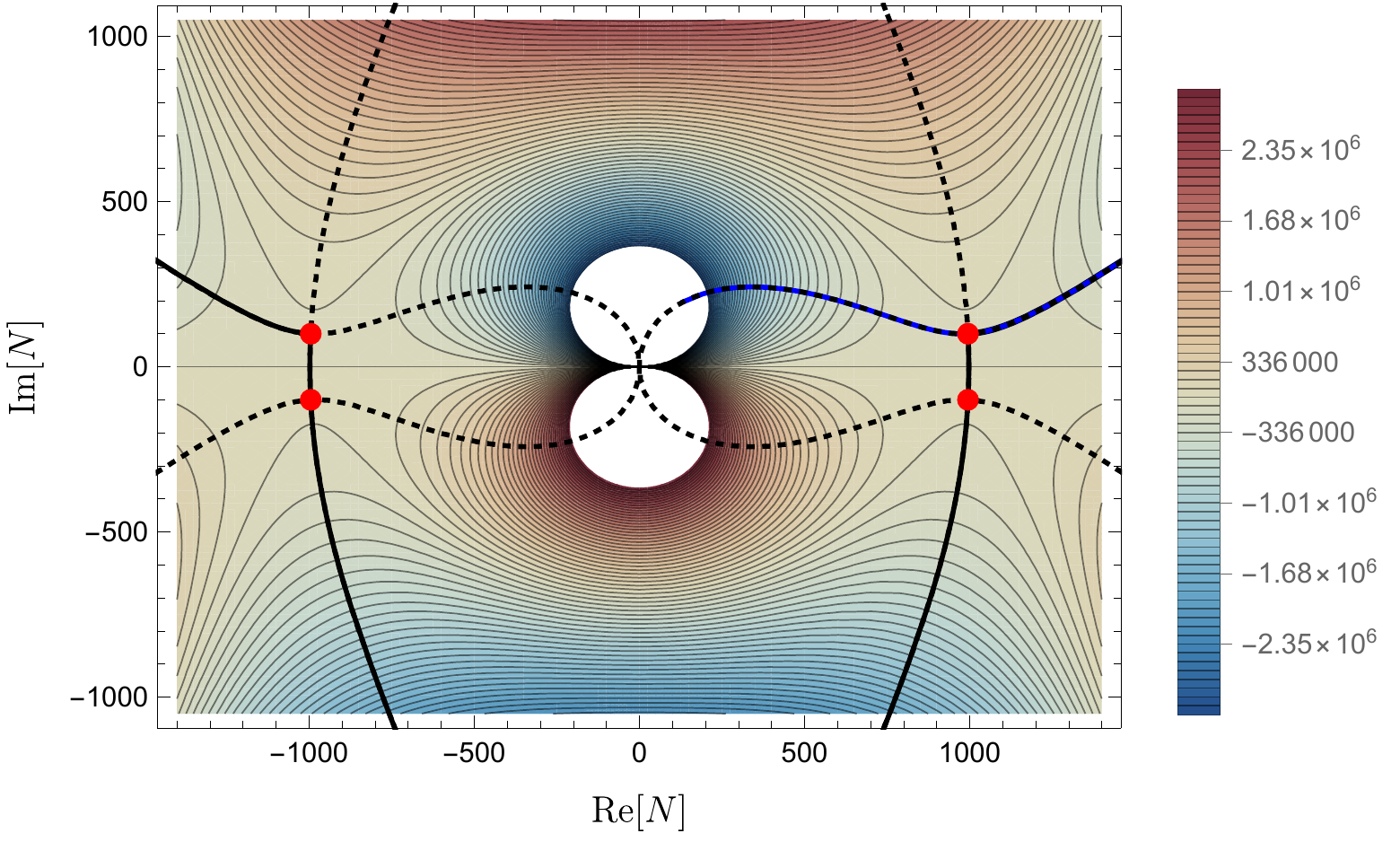}
\caption{
This figure shows $\textrm{Re}\left[iS_{\rm on-shell}^{(0)}[N]\right]$ for 
the on-shell action~\eqref{eq:semi-action1} 
in the complex lapse $N$ plane and we set $H=10^{-1}$, $q_1=10^4$. Steepest descent lines $\cal J_\sigma$
and ascent lines $\cal K_\sigma$ 
are drawn as black dotted lines in the blue and red regions.
The four red dots are the saddle points~\eqref{eq:saddle-points}.
The blue solid line provides 
the tunneling propagator~\eqref{eq:tunneling-propagator} whereas 
the black solid line provides the no-boundary propagator~\eqref{eq:no-boundary-propagator}. }
\label{fig:Picard-Lefschetz1}
\end{figure} 

We plot $\textrm{Re}\left[iS_{\rm on-shell}^{(0)}[N]\right]$ for the on-shell action~\eqref{eq:semi-action1} over the complex plane in Fig.~\ref{fig:Picard-Lefschetz1}.
By utilizing the saddle point approximation, the tunneling propagator can be found with the saddle point with $c_1=c_2=+1$ so that the lapse $N$ integration contour runs along the steepest descent paths known as the Lefschetz thimbles $\cal J_\sigma$ and passes this saddle point. On the other hand, the no-boundary propagator is given by the two saddle points with $c_1=+1,\, c_2=-1$ and $c_1=-1,\, c_2=+1$ meaning $\textrm{Im}[N] < 0$. Naively, the lapse $N$ integration contours that pass these two saddle points correspond to the steepest ascent paths, so these contours must be deformed~\cite{DiazDorronsoro:2017hti}. 
After all, the tunneling and no-boundary propagators at the zeroth-order in perturbation are given by Refs.~\cite{Feldbrugge:2017kzv,DiazDorronsoro:2017hti},
\begin{align}
G^{(0)}_{\rm T}[q_1] &\simeq
\frac{e^{+i\pi\over4}}{2(\Lambda q_1 / 3 - 1)^{1/4}}
e^{-\frac{12 \pi^2}{\hbar \Lambda} \, -i 4\pi^2 \sqrt{\frac{\Lambda}{3}} (q_1 - 3/\Lambda)^{3/2}/\hbar} \quad\quad  (\mbox{tunneling})\label{eq:tunneling-propagator}\,, \\
G^{(0)}_{\rm NB}[q_1] &\simeq
\frac{e^{+12\pi^2/\hbar\Lambda}}{(\Lambda q_1 / 3 - 1)^{1/4}}\cos\left[ \frac{12 \pi^2}{\hbar \Lambda} \left(\frac{ \Lambda q_1}{3} - 1\right)^{3/2} + \frac{3 \pi}{4} \right] \quad (\mbox{no-boundary})\label{eq:no-boundary-propagator}\,,
\end{align}
where $q(\tau=\tau_1)=q_1 > 3/\Lambda$.

\section{Tensor perturbation in General Relativity }
\label{sec:Full-analysis}
In the previous section, the most famous formulations of the wave function $\Psi[g]$ of the universe, the no-boundary proposal~\cite{Hartle:1983ai} and the tunneling proposal~\cite{Vilenkin:1984wp}, were discussed in Lorentzian path integrals. Hereafter, we discuss such wave functions $\Psi[g]$ in the mini-superspace, including a tensor perturbation. 
In GR~\cite{Feldbrugge:2017kzv,DiazDorronsoro:2017hti}, it has been shown that for both wave functions of the universe, the linearized perturbations around a background 
are governed by an inverse Gaussian distribution, which leads to divergent correlation functions, and thus the perturbation is uncontrollable. 
In this section, we clearly show that the problem of the inverse Gaussian wave function for linearized tensor perturbations in the Lorentzian path integral is inevitable as far as one requires the regularity of the on-shell gravitational action.

We can perform the Lorentzian path integral for Eq.~\eqref{G-propagator} in two steps
 if we neglect the back-reaction of the linearized tensor perturbations. 
First, we estimate the path integral with respect to the background $q(\tau)$ and the tensor perturbation $h(\tau)$ by the saddle-point approximation. The result of the first step is given in terms of the classical action of the linear perturbation theory  
\begin{align}\label{eq:full-semi-action}
\begin{split}
S_{\rm GR}[q,h,N] &=
S_{\rm GR}^{(0)}[q,N]
+S_{\rm GR}^{(2)}[h,N] +\mathcal{O}(h^3) \,,
\end{split} 
\end{align}
 and that evaluated on-shell as $\exp \left({{iS_{\rm on-shell}^{(2)}[N]}/{\hbar}}\right)$, up to an overall factor of order unity. Next, we integrate $\exp \left({{iS_{\rm on-shell}^{(2)}[N]}/{\hbar}}\right)$ over the constant lapse function $N$. In this step, we utilize the Picard-Lefschetz method, which complexifies $N$ and 
selects complex integration contours
based on the steepest descent paths known as the Lefschetz thimbles $\cal J_\sigma$. Since $\cal J_\sigma$ ensures the convergence of the integral, 
we can efficiently perform the Lorentzian path integral.

Let us now perform the first step, i.e. the path integral with respect to $q(t)$ and $h(t)$ by the saddle-point approximation. 
We can write down the second-order action for the tensor perturbation in terms of
 the new time coordinate $\tau$ defined in~\eqref{new-time} and the variable $q(\tau)$ defined in~\eqref{def-q},
\begin{align}
\begin{split}\label{action2}
S_{\rm GR}^{(2)}[h,N] =  2\pi^2 \int_{0}^{1} N \mathrm{d}\tau \, 
\left\{ \frac{q^2}{8N^2}\dot{h}^2
-\frac{1}{8}\left[(n^2-3)+6 \right]h^2\right\} \,,
\end{split}
\end{align}
and evaluate the on-shell action $S_{\rm on-shell}^{(2)}[N] $,
\begin{align} \label{eqn:S2onshell}
S_{\rm on-shell}^{(2)}[N] &=
  \frac{\pi^2}{4} \left[ q^2 \frac{h \dot{h}}{N} \right]^1_0\,,
  \end{align}
where we performed the integration by parts for the action~\eqref{action2}
and used the equation of motion for $h(\tau)$.

For convenience, let us introduce $\chi(\tau)=q(\tau) h(\tau) $ and
write the equation of motion for $\chi(\tau)$ as
\begin{align}\label{eq:eom-original}
\frac{\ddot{\chi}}{N^2}  + \left[ \frac{(n^2-3)+6}{q^2} - \frac{1}{N^2} \frac{\ddot{q}}{q}\right]\chi= 0 \,.
\end{align} 
Given the classical solution for the background $q(\tau)=N^2H^2 \tau(\tau-1)+q_1\tau$
which satisfies the boundary condition $q(0)=0$
and $q(1)=q_1$, we have the solution for the above equation (\ref{eq:eom-original}) as
\begin{align}\label{eq:gr-solution}
&\chi(\tau)=\sqrt{ \left(H^2 N^2 \tau (\tau -1)+q_1\tau \right) 
\left(H^4 N^2\tau(\tau -1)  +H^2 q_1 \tau + \alpha \right)}\\
&\times  \Biggl\{ C_1 \left(\frac{H^2 N^2 (\tau -1)+q_1}{\tau }\right)^{\gamma \over2} \sqrt{\frac{H^2 N^2 (2 \tau -1)+\sqrt{H^4 N^4+N^2 \left(-4 \alpha -2 H^2 q_1\right)+q_1^2}+q_1}{H^2 N^2 (2 \tau -1)-\sqrt{H^4 N^4+N^2 \left(-4 \alpha -2 H^2 q_1\right)+q_1^2}+q_1}}\notag \\
&+C_2 \left(\frac{\tau }{H^2 N^2 (\tau -1)+q_1}\right)^{
\gamma \over2}\sqrt{\frac{H^2 N^2 (2 \tau -1)-\sqrt{H^4 N^4+N^2 \left(-4 \alpha -2 H^2 q_1\right)+q_1^2}+q_1}{H^2 N^2 (2 \tau -1)+\sqrt{H^4 N^4+N^2 \left(-4 \alpha -2 H^2 q_1\right)+q_1^2}+q_1}}\Biggr\}\notag ,
\end{align}
where $C_{1,2}$ are constants, and we have defined $\gamma=\sqrt{1-\frac{4\alpha N^2}
{\left(q_1-N^2H^2\right)^2}}$ and $\alpha=(n^2-3)+6$.
In order to estimate the on-shell action, we only need the values of $\chi(\tau)$ and $\dot{\chi}(\tau)$ at the two boundaries $\tau = 0, 1$ (see (\ref{eqn:S2onshell})). 
The on-shell action for the solution~\eqref{eq:gr-solution} is written as
\begin{align}\label{eq:semiclassical-action}
&S_{\rm on-shell}^{(2)}[N]=-\frac{\pi^2 \alpha }{8 N}\Biggl[
C_1^2 q_1^{\gamma} \left(\sqrt{\left(q_1-H^2 N^2\right)^2-4 \alpha  N^2}+H^2 N^2+q_1\right)+C_2^2q_1^{-\gamma}  \\
&\times \left(-\sqrt{\left(q_1-H^2 N^2\right)^2-4 \alpha  N^2}+H^2 N^2+q_1\right) +2 C_1C_2 \left(H^2 N^2+q_1\right)\Biggr]
-{\pi^2\over 4 N}q^2\dot{h}h \Bigr|_{\tau=0}\notag\,.
\end{align}

Near the boundary $\tau = 0$, the solution~\eqref{eq:gr-solution} behaves as 
\begin{equation}\label{eq:gr-solution0}
\chi(\tau)\propto C_1\, F_1[N] \tau^{\frac{1}{2}(1-\gamma)}
+C_2\, F_2[N]\tau^{\frac{1}{2}(1+\gamma)}\quad (\tau \to 0)\, ,
\end{equation}
where $F_1[N]$, $F_2[N]$ are functions of $N$ whose explicit form can be derived from the general solution~\eqref{eq:gr-solution}. From this expression one can show that the contribution of $\tau=0$ to the on-shell action (\ref{eqn:S2onshell}) contains terms of the form $\propto C_1\tau^{-\gamma}$, $\propto C_1C_2$ and $\propto C_2^2\tau^{\gamma}$, and thus is finite if and only if $C_1=0$ (or $C_2=0$) for $\textrm{Re}[\gamma]>0$ (or for $\textrm{Re}[\gamma]<0$, respectively).

Setting either $C_1=0$ (for $\textrm{Re}[\gamma]>0$) or $C_2=0$ (for $\textrm{Re}[\gamma]<0$), and then fixing the remaining integration constant $C_2$ or $C_1$, respectively, by $\chi(1)=q_1h_1$ for the solution~\eqref{eq:gr-solution}, we obtain
\begin{align}\label{eq:gr-semiclassical-action}
S_{\rm on-shell}^{(2)}[N]=\begin{cases} 
-\frac{\pi ^2 q_1h_1^2\alpha \left(\sqrt{\left(q_1-H^2 N^2\right)^2-4 \alpha  N^2}+H^2 N^2+q_1\right)}{8 N \left(\alpha +H^2 q_1\right)}
-{\pi^2q^2\dot{h}h\over 4 N} \Bigr|_{\tau=0}\ 
\textrm{for}\ \textrm{Re}[\gamma]>0 \\ 
-\frac{\pi ^2 q_1h_1^2\alpha \left(-\sqrt{\left(q_1-H^2 N^2\right)^2-4 \alpha  N^2}+H^2 N^2+q_1\right)}{8 N \left(\alpha +H^2 q_1\right)}
-{\pi^2q^2\dot{h}h\over 4 N}\Bigr|_{\tau=0}\ \textrm{for}\ \textrm{Re}[\gamma]<0. \end{cases}
\end{align}

\begin{figure}
	\subfigure[$H=10^{-1}$, $q_1=10^4$, $h_1= 0.01$, $n=0$]{%
		\includegraphics[clip, width=0.5\columnwidth]
		{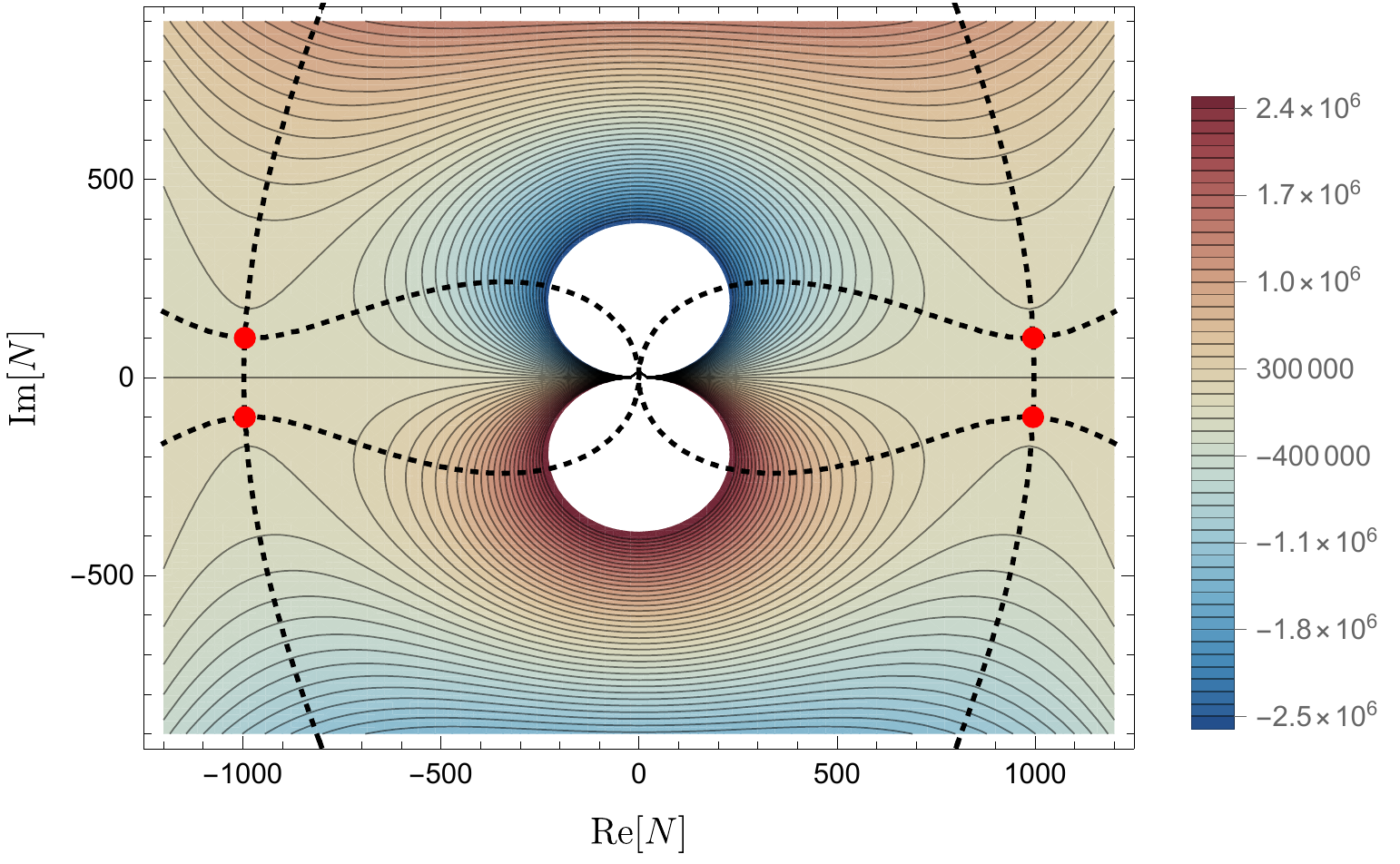}}%
	\subfigure[$H=10^{-1}$, $q_1=3.0\times 10^2$, $h_1= 0.01$, $n=0$]{%
		\includegraphics[clip, width=0.5\columnwidth]
		{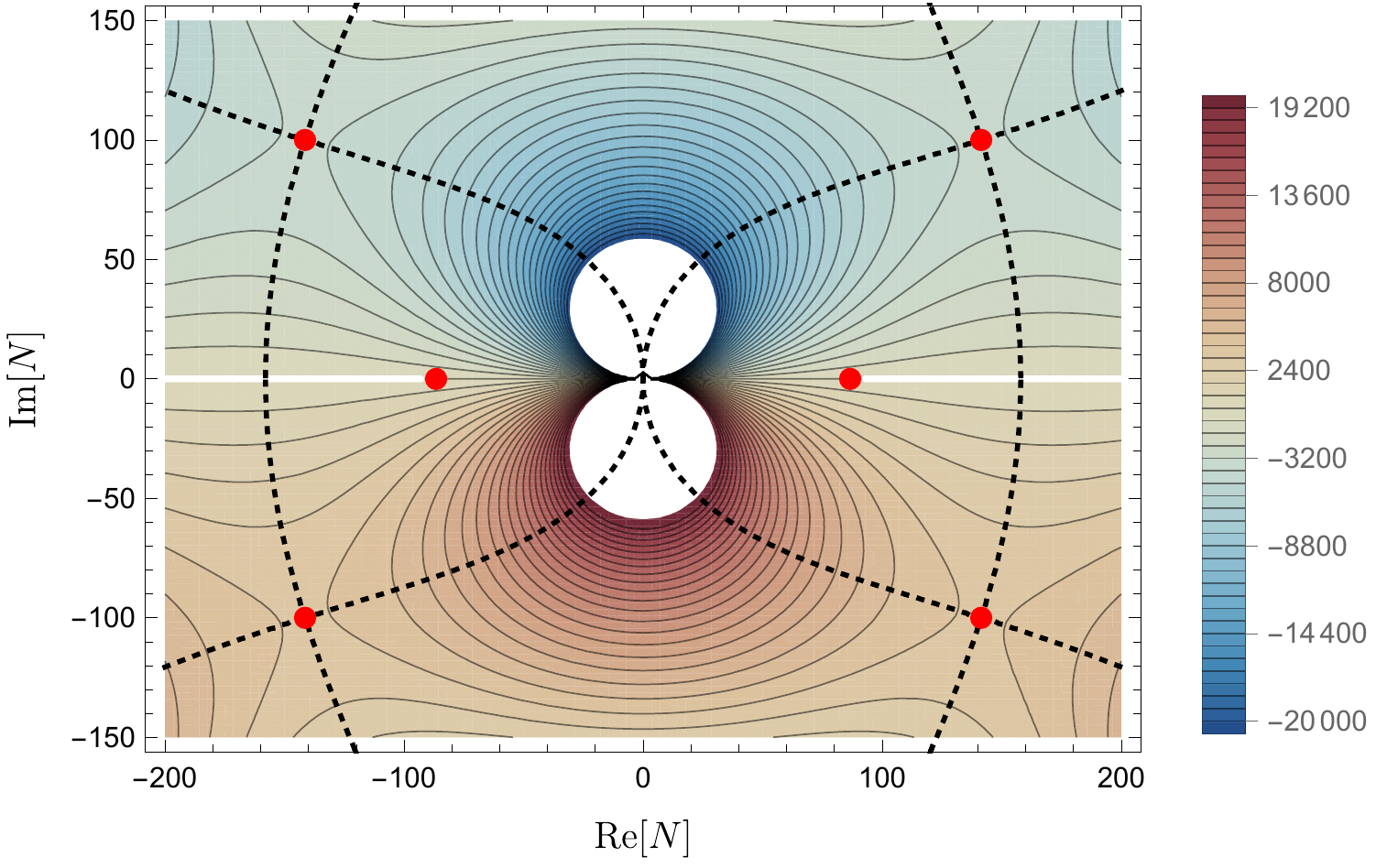}}\\
         \subfigure[$H=10^{-2}$, $q_1=10^4$, $h_1= 0.01$, $n=0$]{%
		\includegraphics[clip, width=0.5\columnwidth]
		{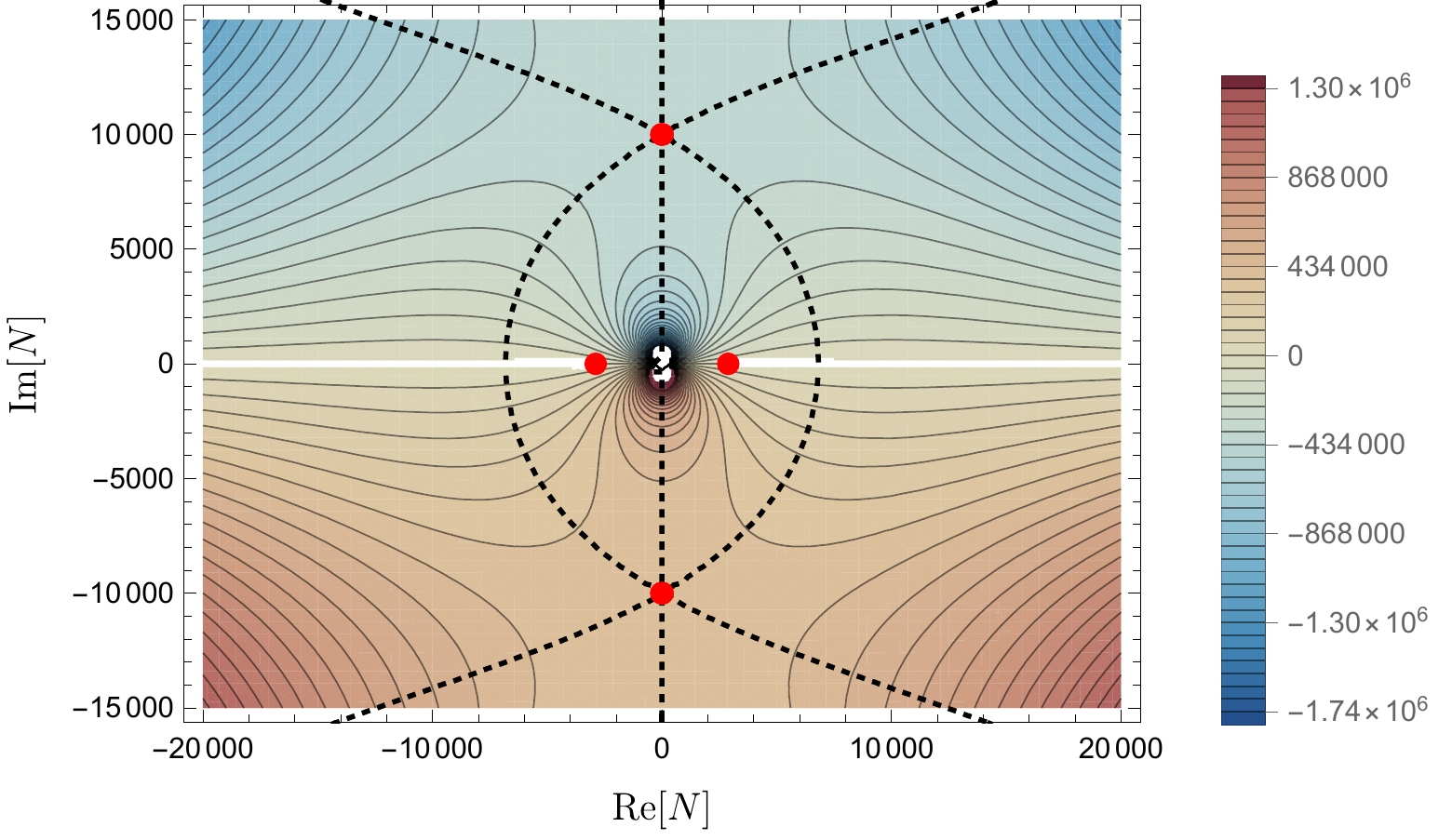}}%
          \subfigure[$H=10^{-2}$, $q_1=0.9\times 10^4$, $h_1= 0.01$, $n=0$]{%
		\includegraphics[clip, width=0.5\columnwidth]
		{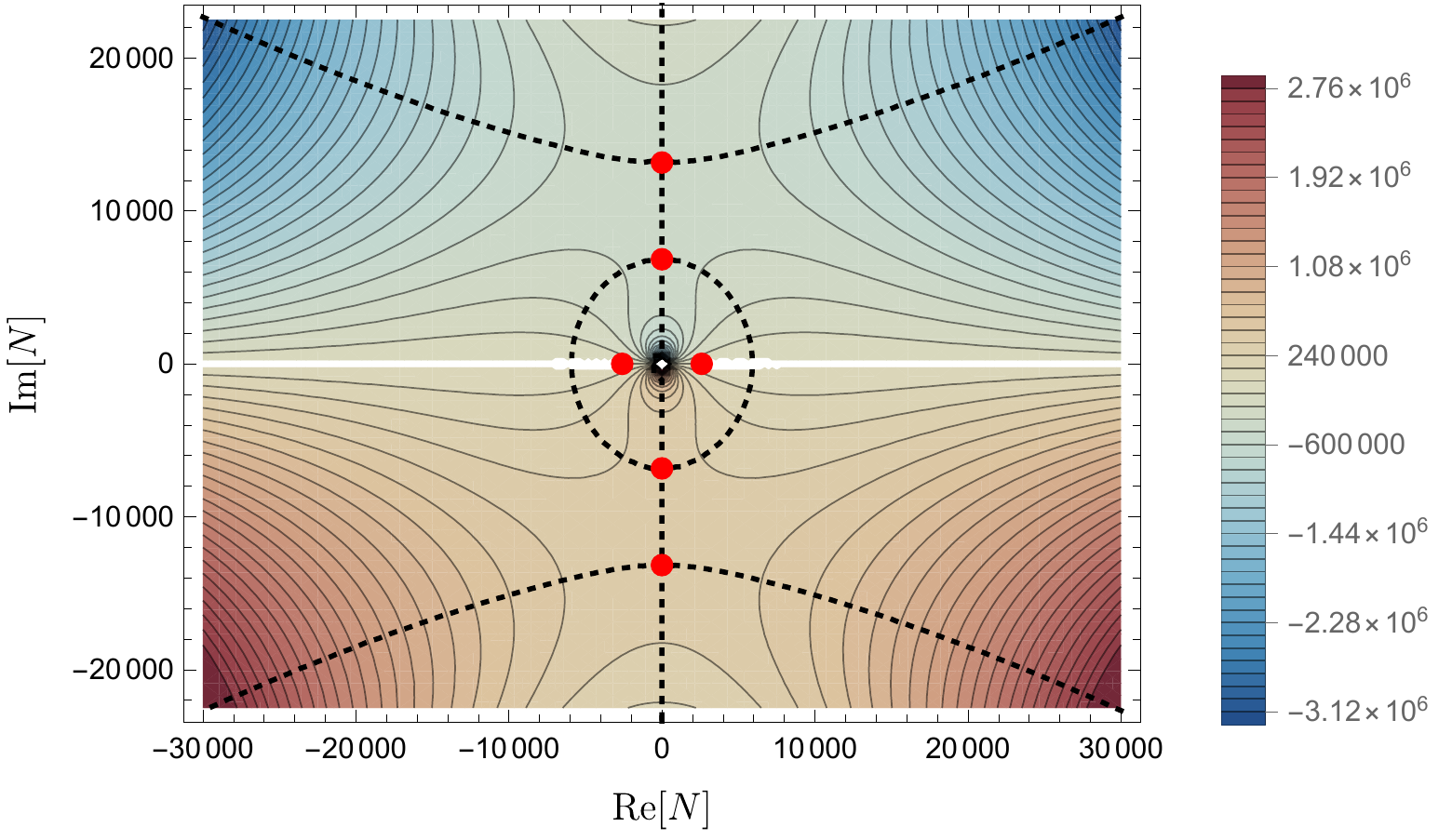}}%
	\caption{These figures show 
	$\textrm{Re}\left[iS_{\rm on-shell}[N]\right]$ for 
the on-shell action of~\eqref{eq:full-semi-action} 
in the complex lapse plane. Steepest descent lines $\cal J_\sigma$
and ascent lines $\cal K_\sigma$ 
are drawn as black dotted lines in the blue and red regions.
White lines express  the branch cut
on the complex $N$-plane.}
	\label{fig:Picard-Lefschetz2}
\end{figure}

Hereafter, we ignore the back-reaction of the tensor perturbations to the background. In particular, we do not take into account a possible shift of the position of the saddle points in the complex-$N$ plane due to the tensor perturbation and simply evaluate $S_{\rm on-shell}^{(2)}[N]$ for $\chi(\tau)$ at the background saddle points in the complex-$N$ plane. 
This would be a good approximation
as long as the back-reaction of the perturbations is negligible.
In Fig.~\ref{fig:Picard-Lefschetz2} we plot $\textrm{Re}\left[iS_{\rm on-shell}[N]\right]$ for the on-shell action of~\eqref{eq:full-semi-action} with various parameters, 
and also confirm the background saddle-points do not change significantly for the small $h$.
For simplicity, we consider $\textrm{Re}[N] > 0$ so that there exist 
two background saddle-points for $N \in (0 , \infty)$, 
\begin{align}\label{eq:tunneling-saddle}
N_\textrm{T}&=\frac{1}{H^2}
\left[i+\left(q_{1}H^2-1\right)^{1/2}\right],\\ \label{eq:no-boundary-saddle}
N_\textrm{HH}&=-\frac{1}{H^2}
\left[i-\left(q_{1}H^2-1\right)^{1/2}\right].
\end{align}

Taking the tunneling saddle point 
$N_\textrm{T}$~\eqref{eq:tunneling-saddle} for 
the on-shell action~\eqref{eq:gr-semiclassical-action} with $C_1=0$ (for $\textrm{Re}[N] > 0$),
we have
\begin{align}\label{eq:gr-semiclassical-action-ts}
\frac{i}{\hbar} S_{\rm on-shell}^{(2)}[N_\textrm{T}] 
&=+\frac{\pi ^2 q_1h_1^2 \alpha  \left(\sqrt{\alpha +1}-i\sqrt{H^2q_1-1}\right)}
{4\hbar \left(\alpha +H^2 q_1\right)} \notag\\
&\approx \frac{\pi^2\alpha^{3/2}}{4\hbar H^2}\left[ 1 - i q_1^{1/2}\alpha^{-1/2} + \cdots \right] h_1^2\,,
\end{align}
where we took the super-horizon mode ($n\ll q_1^{1/2}H$)
and $q_1\gg 1/H^2$ in the last expression.
The real part of (\ref{eq:gr-semiclassical-action-ts}) is positive and thus leads to an inverse Gaussian distribution for the tensor perturbations, meaning that the perturbations are out of control. On the other hand, when we consider the on-shell action~\eqref{eq:semiclassical-action} at the no-boundary saddle point $N_\textrm{HH}$, we have a Gaussian distribution for the tensor perturbations. 
Although the contribution of the no-boundary saddle points to the correlation functions of linearized perturbations is finite, the lapse $N$ integration contours must be deformed~\cite{DiazDorronsoro:2017hti} and pass through the tunneling saddle point $N_\textrm{T}$ and the branch cuts which again lead to divergent correlation functions of linearized perturbations~\cite{Feldbrugge:2017mbc}.
Therefore, the no-boundary
propagator~\eqref{eq:no-boundary-propagator} is also inconsistent once taking into account the tensor perturbations.
This is a brief overview of the problems of linearized 
perturbations for the no-boundary~\cite{Hartle:1983ai} and
tunneling proposals~\cite{Vilenkin:1984wp} in Lorentzian path integral in GR.
\footnote{
In Refs.~\cite{DiTucci:2019dji,DiTucci:2019bui} 
the authors proposed different boundary conditions for the background to rescue the no-boundary propagator~\eqref{eq:no-boundary-propagator} but 
such modification abandons the quantum creation of the 
universe from nothing $q(0)=0$.
Since this modification also does not rescue the tunneling proposal, 
we do not consider such different boundary conditions for the background in this paper.}

As a counterargument to these difficulties, in Ref.~\cite{Halliwell:2018ejl} 
the authors abandon the (Lorentzian) path integral 
and suggest that the semiclassical no-boundary wave function 
should be given by the assumptions of saddle point uniqueness 
(see also Refs.~\cite{deAlwis:2018sec,Matsui:2021oio} for the related works).
Furthermore, the recently proposed allowability criterion of~\cite{Kontsevich:2021dmb} (see also Ref.~\cite{Witten:2021nzp,Lehners:2021mah,Jonas:2022njf}) 
restricts complex metrics of the gravitational path integral to 
require p-form gauge theories to be well defined. 
Ref.~\cite{Lehners:2021mah} shows that the Lorentzian path integral formulation 
intrinsically conflicts with this allowability criterion and
the steepest descent contours for the no-boundary lapse integral
enter into non-allowable regions. It seems that 
the meaning of Lorentzian path integral is not entirely clear,
and there are no solutions to reflect the classical behavior of the system
in the path integral for two-dimensional indefinite oscillator model~\cite{Kiefer:1989va,Kiefer:1990ms}.
Thus, further investigations might be necessary to obtain 
a consistent path integral formulation of quantum gravity.
We shall not discuss these subtle open issues further. 
In the next section, we shall instead consider the inverse Gaussian problem of tensor perturbations with the modified dispersion relation in light of trans-Planckian physics.

\section{Trans-Planckian physics and modified dispersion relation}
\label{sec:TPP}

In this section, we will discuss the perturbation problem in 
Lorentzian quantum cosmology by assuming modified dispersion relations motivated by the trans-Planckian physics~\cite{Martin:2000xs,Brandenberger:2000wr,
Niemeyer:2000eh,Martin:2002kt,Ashoorioon:2004vm,Ashoorioon:2011eg}. 
The modified dispersion relation can take the form $\omega^2 = {\cal F}\, ({k_{\rm phys}})$~\cite{Brandenberger:2012aj}, where $k_{\rm phys}=\alpha^{1\over 2}/{q^{1\over 2}}$ is the physical wavenumber and $\alpha=(n^2-3)+6$. Since the physical momentum diverges at the classical big-bang singularity $q \to 0$, the dispersion relation would drastically change. For instance, we can introduce the modified dispersion relation, including the trans-Planckian cutoff~\cite{Niemeyer:2000eh}, 
\begin{align}\label{Modified-dispersion-cutoff}
{\cal F}\, (k_{\rm phys})=\begin{cases} 
k_{\rm phys}^2 \quad \textrm{for}\quad k_{\rm phys}^2\ll {\cal M}_{\rm UV}^2 \\ 
{\cal M}_{\rm UV}^2\quad \textrm{for}\quad k_{\rm phys}^2\gg {\cal M}_{\rm UV}^2, \end{cases}
\end{align}
where ${\cal M}_{\rm UV}$ is the trans-Planckian cutoff scale. 
In the limit $q\to 0$, correspondingly $k_{\rm phys} \to \infty$, the dispersion relation for all modes is modified from that in GR. 
A concrete example of the trans-Planckian cutoff is the Unruh dispersion relation with an arbitrary constant $b$~\cite{Unruh:1994je},
\begin{align}\label{Modified-dispersion-Unruh}
{\cal F}\, (k_{\rm phys}) 
={\cal M}_{\rm UV}^2
\tanh^{2/b}\left[\left(\frac{k_{\rm phys}^2}{{\cal M}_{\rm UV}^2}\right)^{b\over2}\right],
\end{align}
which satisfies (\ref{Modified-dispersion-cutoff}). 
On the other hand, in Ref.~\cite{Corley:1996ar,Corley:1997pr}, 
the authors introduced the following modified 
dispersion relation in the context of black holes physics,
\begin{align}\label{Modified-dispersion-Corley/Jacobson}
{\cal F}\, (k_{\rm phys})
= k_{\rm phys}^2+\frac{k_{\rm phys}^4}{{\cal M}_{\rm UV}^2} \,.
\end{align}
The following more general expression was considered by Ref.~\cite{Martin:2000xs},
\begin{align}\label{Modified-dispersion-general}
{\cal F}\, (k_{\rm phys})
= k_{\rm phys}^2
+k_{\rm phys}^2\sum^{p}_{j=1}b_j\left(\frac{k_{\rm phys}^2}{{\cal M}_{\rm UV}^2}\right)^{j}, 
\end{align}
where the right-hand side should be non-negative for all $k_{\rm phys}^2 \geq 0$ to avoid instability.
The expression~\eqref{Modified-dispersion-general} includes the modified dispersion relation introduced by higher-dimensional operators in higher-curvature theories of gravity such as Ho\v{r}ava-Lifshitz gravity~\cite{Horava:2009uw}. Therefore, our analysis of linear perturbations with this dispersion relation is expected to be applicable to the analysis of those theories if the background dynamics are also properly modified. We will consider the Unruh dispersion relation~\cite{Unruh:1994je} and a special case of the generalized Corley-Jacobson dispersion relation~\cite{Corley:1996ar,Corley:1997pr} as examples of the modified dispersion relations. Focusing on the contribution of the $\tau\to 0$ boundary to $S^{(2)}[N]$, we discuss whether the perturbative on-shell action can be rendered regular by these dispersion relations for $h$.

Now, let us consider the second-order action $S^{(2)}[h,N]$ for the tensor perturbation $h$ with the dispersion relation $\omega^2 = {\cal F}\, (k_{\rm phys})$, where we defined $k_{\rm phys} = \alpha^{1\over 2}/q^{1\over 2}$ and $\alpha=(n^2-3)+6$, and rewrite it for $\chi=qh$ as follows:
\begin{align}
 S^{(2)}[h,N]&= 2\pi^2 \int_{0}^{1} N \mathrm{d}\tau \, \left[ \frac{q^2}{8N^2} \dot{h}^2 -\frac{q }{8}{\cal F}(k_{\rm phys}) h^2 \right]+S^{(2)}_{B} \notag\\
 &= \frac{\pi^2}{4} \int_{0}^{1} N \mathrm{d}\tau \, \left[ \frac{1}{N^2}\left(\dot{\chi}^2-2\frac{\dot{\chi}\chi\dot{q}}{q}+\frac{\chi^2\dot{q}^2}{q^2}\right)
 -{\cal F}(k_{\rm phys})\frac{\chi^2}{q} \right]+S^{(2)}_{B}\,,
\end{align}
where we added possible boundary contributions $S^{(2)}_{B}$ of the perturbations localised on the hypersurfaces at $\tau=0,1$.
We note that for tensor perturbations, the boundary terms of the background and perturbations are interrelated, and modifying the boundary terms of the tensor perturbations $S^{(2)}_{B}$ also modifies the boundary terms of the background $S^{(0)}_{B}$. 
Hence, it is not straightforward to introduce a suitable boundary term in a consistent manner. Therefore, we do not introduce any boundary terms of the tensor perturbations $S^{(2)}_{B}$ and assume the Dirichlet boundary conditions, $\chi\mid^{t=1}_{t=0} = \textrm{fixed.}$.

Since the above action depends on $\chi$ and $\dot{\chi}$, 
the variation of the action in terms of $\chi$ is given by 
\begin{align}
\begin{split}
\delta S^{(2)}[ \chi,N]&= \frac{\pi^2}{4}\int_0^1 Nd\tau \, \Bigl[
\frac{\partial S^{(2)}}{\partial \chi}\delta \chi+ \frac{\partial S^{(2)}}{\partial \dot{\chi}}\delta \dot{\chi}\Bigr]
\\
&=\frac{\pi^2}{4}\int_0^1 Nd\tau \, \Bigl[
\left\{
\frac{1}{N^2}\left(-2\frac{\dot{\chi}\dot{q}}{q}+2\frac{\chi\dot{q}^2}{q^2}\right)-
{\cal F}(k_{\rm phys})\frac{2\chi}{q}\right\}
\delta \chi \\
&\quad +\left\{
\frac{1}{N^2}\left(2\dot{\chi}-2\frac{\chi\dot{q}}{q}\right)\right\}
\delta \dot{\chi}\Bigr]
\\
&=-\frac{\pi^2}{2}\int_0^1 Nd\tau \, \Bigl[
\left\{\frac{1}{N} \partial_\tau \left( \frac{\dot{\chi}}{N} \right)
 +\left[ \frac{{\cal F}(k_{\rm phys})}{q}- \frac{1}{qN} \partial_\tau \left( \frac{\dot{q}}{N} \right) \right] \chi \right\}
\delta \chi\Bigr]\\
&\quad+ \frac{\pi^2}{2N}\left(\dot{\chi}-\frac{\chi\dot{q}}{q}\right)\delta \chi\mid^{t=1}_{t=0}
\label{variation-perturbations}
\end{split}
\end{align}
It should be noted that the last term will disappear due to the Dirichlet boundary conditions $\chi\mid^{t=1}_{t=0} = \textrm{fixed}$.
From this action, the equation of motion for $\chi(\tau)$ reads
 \begin{align}\label{eq:eom2}
 \frac{1}{N} \partial_\tau \left( \frac{\dot{\chi}}{N} \right)
 +\left[ \frac{{\cal F}(k_{\rm phys})}{q}- \frac{1}{qN} \partial_\tau \left( \frac{\dot{q}}{N} \right) \right] \chi = 0 \,.
\end{align}
One can evaluate the on-shell value of $S^{(2)}[h,N]$ as
 \begin{align} \label{eqn:onshellaction-boundarydata}
 S_{\rm on-shell}^{(2)}[N]
 &=  \frac{\pi^2}{4} \left[ q^2 \frac{h \dot{h}}{N} \right]^1_0 \,.
\end{align}
This formal expression of the on-shell action $S_{\rm on-shell}^{(2)}[N]$ in terms of the boundary data is exactly the same as that in GR irrespectively of the functional form of ${\cal F} (k_{\rm phys})$, while the boundary data depends on the equations of motion in bulk (as well as the boundary condition) and thus on the choice of ${\cal F} (k_{\rm phys})$. We will use this expression to evaluate the on-shell action $S_{\rm on-shell}^{(2)}[N]$ with the modified dispersion relation.

\subsection{Generalized Corley-Jacobson dispersion relation}

In this subsection, we consider the generalized Corley-Jacobson 
dispersion relation and discuss the issue of the inverse Gaussian wave function 
for tensor perturbations with this dispersion relation~\eqref{Modified-dispersion-general}.
We consider the equation of motion~\eqref{eq:eom2} with 
the generalized dispersion relation~\eqref{Modified-dispersion-general},
 \begin{align}
 \frac{1}{N} \partial_\tau \left( \frac{\dot{\chi}}{N} \right)
+\left\{ \frac{\alpha}{q^2}
+\frac{\alpha}{q^2}\sum^{p}_{j=1}b_j\left(\frac{\alpha}{q
{\cal M}_{\rm UV}^2}\right)^{j}+ \frac{1}{qN} \partial_\tau \left( \frac{\dot{q}}{N} \right) \right\} \chi = 0 \,.
\end{align}
To simplify the analysis, we assume that the dispersion relation~\eqref{Modified-dispersion-general} only contains the last term of the sum and consider $p=2$. 
Thus, we have
\begin{align}\label{eq:eom-jc}
\frac{\ddot{\chi}}{N^2}  + \left\{ \frac{\alpha}{q^2}
\left[1+b_2\left(\frac{\alpha}{q
{\cal M}_{\rm UV}^2}\right)^{2}\, \right]- \frac{1}{N^2} \frac{\ddot{q}}{q}\right\}\chi= 0 \,.
\end{align}

First, we consider the ultraviolet (UV) regime $\alpha/q \gg {\cal M}_{\rm UV}^2$, and simply seek the solutions for the tensor perturbations and calculate the contribution of the UV boundary, i.e. the classical big-bang singularity ($\tau=0$), to the on-shell action. We will consider the following equation of motion in the UV regime,
\begin{align}
\frac{\ddot{\chi}}{N^2}  + \left\{ \frac{\alpha}{q^2}
b_2\left(\frac{\alpha}{q
{\cal M}_{\rm UV}^2}\right)^{2}
- \frac{1}{N^2} \frac{\ddot{q}}{q}\right\}\chi= 0 \,.
\end{align}
By using the background solution $q(\tau)=N^2H^2 \tau(\tau-1)+q_1\tau$, 
the equation of motion in the UV is rewritten as
\begin{align}
\frac{\ddot{\chi}}{N^2}  + \left\{ \frac{\alpha \beta }{\left(H^2 N^2 (\tau -1) \tau +q_1 \tau \right)^4}-\frac{2 H^2}{H^2 N^2 (\tau -1) \tau +q_1 \tau }\right\}\chi= 0 \,,
\end{align}
where $\beta=b_2\alpha^2/{\cal M}_{\rm UV}^4$, and the general solution is 
\begin{align}\label{eq:gcj-solution}
\chi(\tau)&=C_3\left(N^{2} H^{2} \left(\tau-1\right)+q_1\right)^{\zeta_1}\tau^{\, \zeta_2} 
e^{-\frac{\sqrt{-\alpha\beta}\, N \left(N^{2}H^{2}\left(2\tau-1\right) +q_1\right)}
{\left(N^{2} H^{2}-q_1\right)^{2}\left(N^{2}H^{2}
\left(\tau-1\right)+q_1\right)\tau}} \notag\\
&+C_4 \left(N^{2} H^{2} \left(\tau-1\right)+q_1\right)^{\zeta_2}\tau^{\, \zeta_1}
e^{+\frac{\sqrt{-\alpha\beta}\, N \left(N^{2}H^{2} \left(2\tau-1\right) +q_1\right)}{\left(N^{2} H^{2}-q_1\right)^{2}\left(N^{2} H^{2}\left(\tau-1\right)+q_1\right)\tau}}, 
\end{align}
where $C_{3,4}$ are constants and we define $\zeta_{1,2}$ as, 
\begin{align}
\zeta_1=1 - 2 {\frac{N^{3} H^{2} \sqrt{-\alpha\beta}}{\left(N^{2} H^{2}-q_1\right)^{3}}}, \quad 
\zeta_2=1 + 2 \frac{N^{3} H^{2} \sqrt{-\alpha\beta}}{\left(N^{2} H^{2}-q_1\right)^{3}}.
\end{align}
As shown in (\ref{eqn:onshellaction-boundarydata}), the on-shell action has two contributions, one from the UV ($\tau=0$) and the other from the IR ($\tau=1$). In order to estimate the UV contribution, we approximate the above solution~\eqref{eq:gcj-solution} near the boundary $\tau = 0$, 
\begin{equation}\label{eq:gcj-solution0}
\chi(\tau)\propto C_3\, F_3[\tau,N] e^{-{\lambda\over \tau}}
+C_4\, F_4[\tau,N]e^{+{\lambda\over \tau}}\, ,
\end{equation}
where $F_3[\tau,N]$ and $F_4[\tau,N]$ are polynomial functions of $\tau$ whose coefficients depend on $N$, and $\lambda=\sqrt{-\alpha \beta}N/(q_1-H^2N^2)^{2}$. It is clear that the UV ($\tau=0$) contribution to the on-shell action with $C_3= 0$ vanishes for $\textrm{Re}[\lambda]<0$ whereas that with $C_4= 0$ vanishes for $\textrm{Re}[\lambda]>0$. Other choices lead to a divergent on-shell action. Hereafter, we adopt the choices that avoid a divergent on-shell action ($C_3= 0$ for $\textrm{Re}[\lambda]<0$ or $C_4= 0$ for $\textrm{Re}[\lambda]>0$) and, as a result, the UV ($\tau=0$) contribution is zero.

For modes satisfying $\beta\gg q_1^2$ ($\alpha/q_1\gg {\cal M}_{\rm UV}^2$ for $b_1=\mathcal{O}(1)$), we can evaluate not only the UV ($\tau=0$) contribution but also the IR ($\tau=1$) one to the on-shell action by using the solution~\eqref{eq:gcj-solution},
\begin{align}
S_{\rm on-shell}^{(2)}[N]&= \left\{ 
 \begin{array}{ll}
  -\frac{\pi^2 \sqrt{-\alpha \beta }}{4}\cdot C_4^2\cdot q_1^{-\frac{4 \sqrt{-\alpha \beta } H^2 N^3}{\left(H^2 N^2-q_1\right)^3}} e^{-\frac{2 \sqrt{-\alpha \beta } N \left(H^2 N^2+q_1\right)}{q_1 \left(q_1-H^2 N^2\right)^2}} & (\textrm{Re}[\lambda]<0)\\
\frac{\pi^2 \sqrt{-\alpha \beta }}{4}\cdot C_3^2\cdot q_1^{\frac{4 \sqrt{-\alpha \beta } H^2 N^3}{\left(H^2 N^2-q_1\right)^3}} e^{+\frac{2 \sqrt{-\alpha\beta } N \left(H^2 N^2+q_1\right)}{q_1 \left(q_1-H^2 N^2\right)^2}} & (\textrm{Re}[\lambda]>0)
 \end{array}
 \right. \,.
\end{align}
As explained above, to avoid divergences of the on-shell action, we have supposed that $C_3=0$ for $\textrm{Re}[\lambda]<0$ and that $C_4=0$ for $\textrm{Re}[\lambda]>0$. By imposing $\chi(1)=q_1h_1$ to normalize the overall factor for the solution~\eqref{eq:gcj-solution} with $C_3=0$ or $C_4=0$, we obtain
\begin{align}\label{eq:gcj-semiclassical-action}
S_{\rm on-shell}^{(2)}[N]=\begin{cases} 
-\frac{\pi ^2 }{4}\sqrt{-\alpha \beta }h_1^2
\quad
\textrm{for}\ \ \textrm{Re}[\lambda]<0 \\ 
+\frac{\pi ^2 }{4}\sqrt{-\alpha \beta }h_1^2
\quad \textrm{for}\ \ \textrm{Re}[\lambda]>0, \end{cases}
\end{align}
where different points from the GR case in the previous section are that the behavior of the tensor perturbations depends on the background saddle point $N_s$ only through the sign of $\textrm{Re}[\lambda]$. 
Indeed, we obtain inverse Gaussian or Gaussian distribution for the tensor perturbations as
\begin{align} \label{eqn:p=2inverseGorG}
\frac{i}{\hbar} S_{\rm on-shell}^{(2)}[N]=\begin{cases} 
+\frac{\pi ^2 }{4\hbar}{\alpha^{3\over2}b_2^{1\over2}\over{\cal M}_{\rm UV}^2}h_1^2
\quad \textrm{for}\ \ \textrm{Re}[\lambda]<0 \\ 
-\frac{\pi ^2 }{4\hbar}{\alpha^{3\over2}b_2^{1\over2}\over{\cal M}_{\rm UV}^2}h_1^2
\quad \textrm{for}\ \ \textrm{Re}[\lambda]>0\,. \end{cases}
\end{align}

We note that $S_{\rm on-shell}^{(2)}[N] $ depends on the lapse function $N$ only through the sign of $\textrm{Re}[\lambda]$. For instance, taking the tunneling saddle point $N_\textrm{T}$ 
we have
\begin{align}\label{eq:tunneling-gcj-relation}
\lambda[N_\textrm{T}]=\frac{\sqrt{-\alpha\beta}N_\textrm{T}}{
(q_1-H^2N_\textrm{T}^2)^{2}}=
-\frac{\alpha^{3\over2}b_2^{1\over2}}
{4q_1{\cal M}_{\rm UV}^2}\left(1+i\sqrt{q_1H^2-1}\right),
\end{align}
which means $\textrm{Re}[\lambda]<0$.
Thus, we must set $C_3= 0$ and, as a result, we obtain the 
inverse Gaussian wave function for the tunneling proposal,
\begin{equation}
G[q_1, h_1] 
= G^{(0)}[q_1]\cdot \exp \left[{+\frac{\pi ^2 }{4\hbar}{\alpha^{3\over2}
b_2^{1\over2}\over{\cal M}_{\rm UV}^2}h_1^2}\right],
\end{equation}
meaning that the tensor perturbations are out of control.
In contrast, the no-boundary saddle point $N_\textrm{H}$ takes $\textrm{Re}[\lambda]>0$
and leads to the Gaussian distribution for the tensor perturbations.
However, as previously discussed in GR,
the integration contours in the complex $N$ plane must pass through the tunneling saddle point $N_\textrm{T}$ even for the no-boundary proposal~\cite{Feldbrugge:2017mbc}.
Hence, even if the dispersion relation is modified 
as the generalized Corley-Jacobson dispersion 
relation~\eqref{Modified-dispersion-general} with $p=2$, 
the tensor perturbation is out of control for UV modes with 
$\beta\gg q_1^2$ ($\alpha/q \gg {\cal M}_{\rm UV}^2$ for $b_1=\mathcal{O}(1)$).
Although we have only obtained analytical solutions for $p=2$,
and make no analytical estimates for $p\neq2$, 
we numerically confirmed a similar behavior for $p\neq2$.

\begin{figure}
	\subfigure[$H=10^{-1}$, $q_1=10^4$, ${\cal M}_{\rm UV}=10$]{%
		\includegraphics[clip, width=0.48\columnwidth]
		{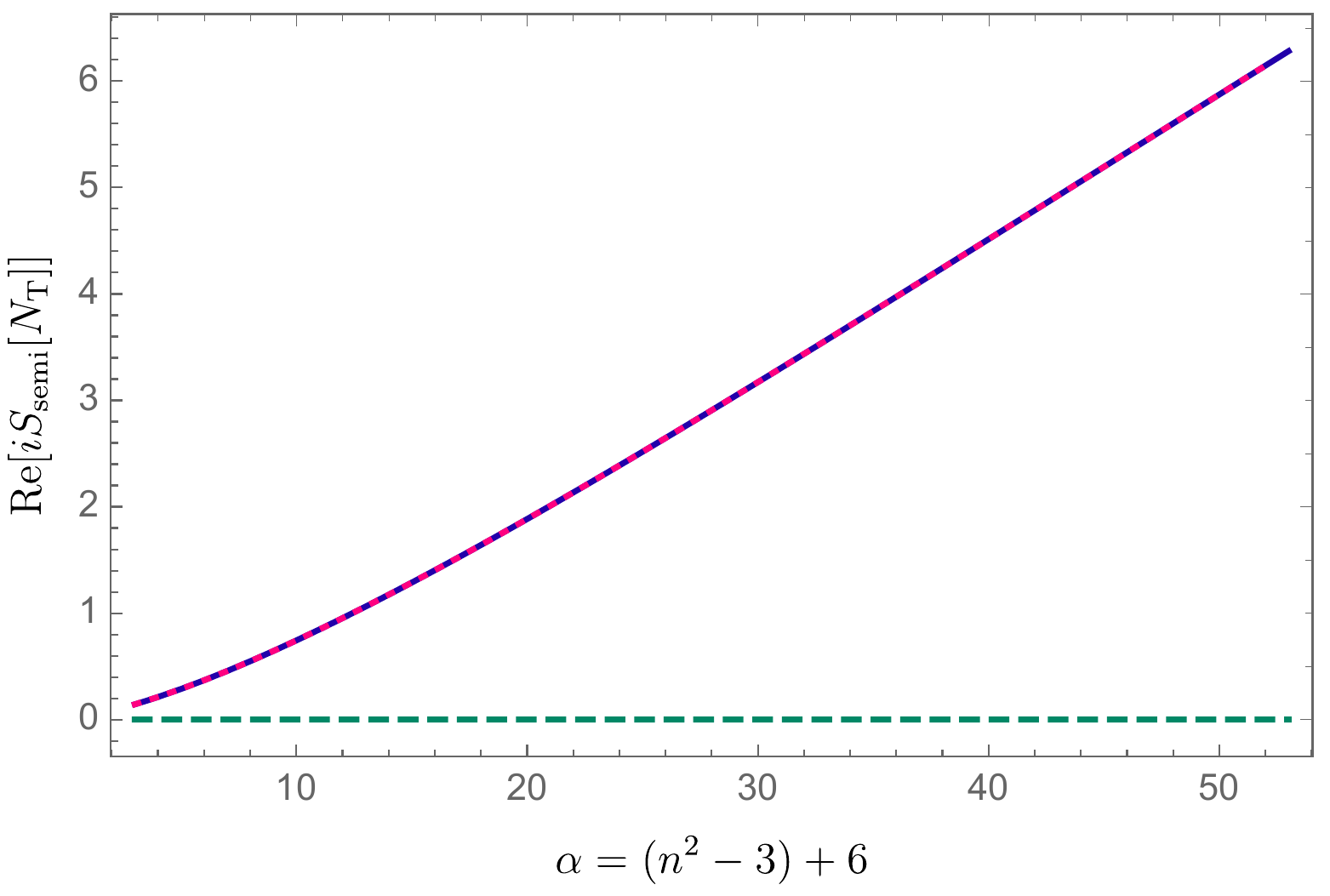}}%
	\subfigure[$H=10^{-2}$, $q_1=10^4$, ${\cal M}_{\rm UV}=10$]{%
		\includegraphics[clip, width=0.48\columnwidth]
		{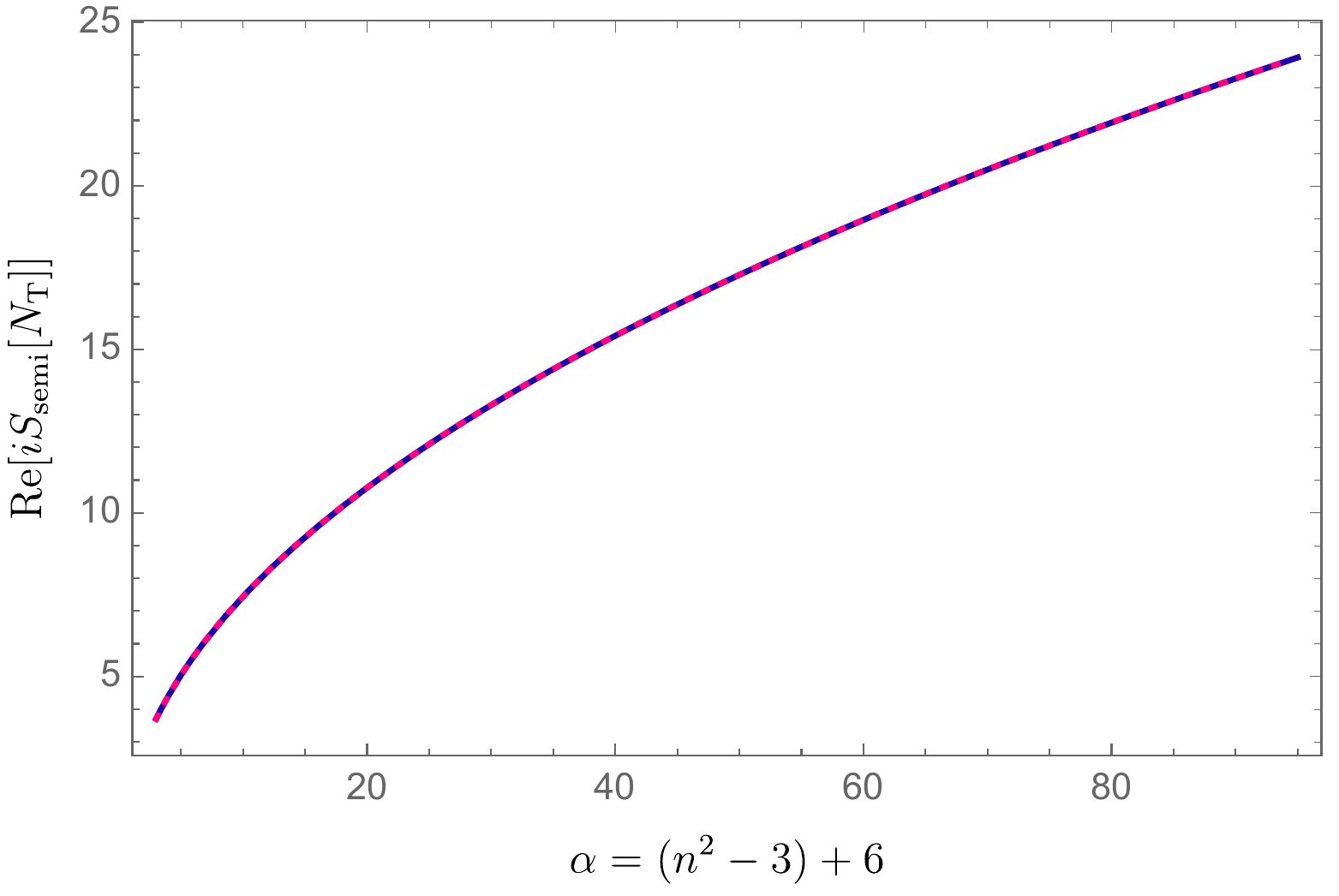}}\\
         \subfigure[$H=10^{-3}$, $q_1=10^3$, ${\cal M}_{\rm UV}=1$]{%
		\includegraphics[clip, width=0.48\columnwidth]
		{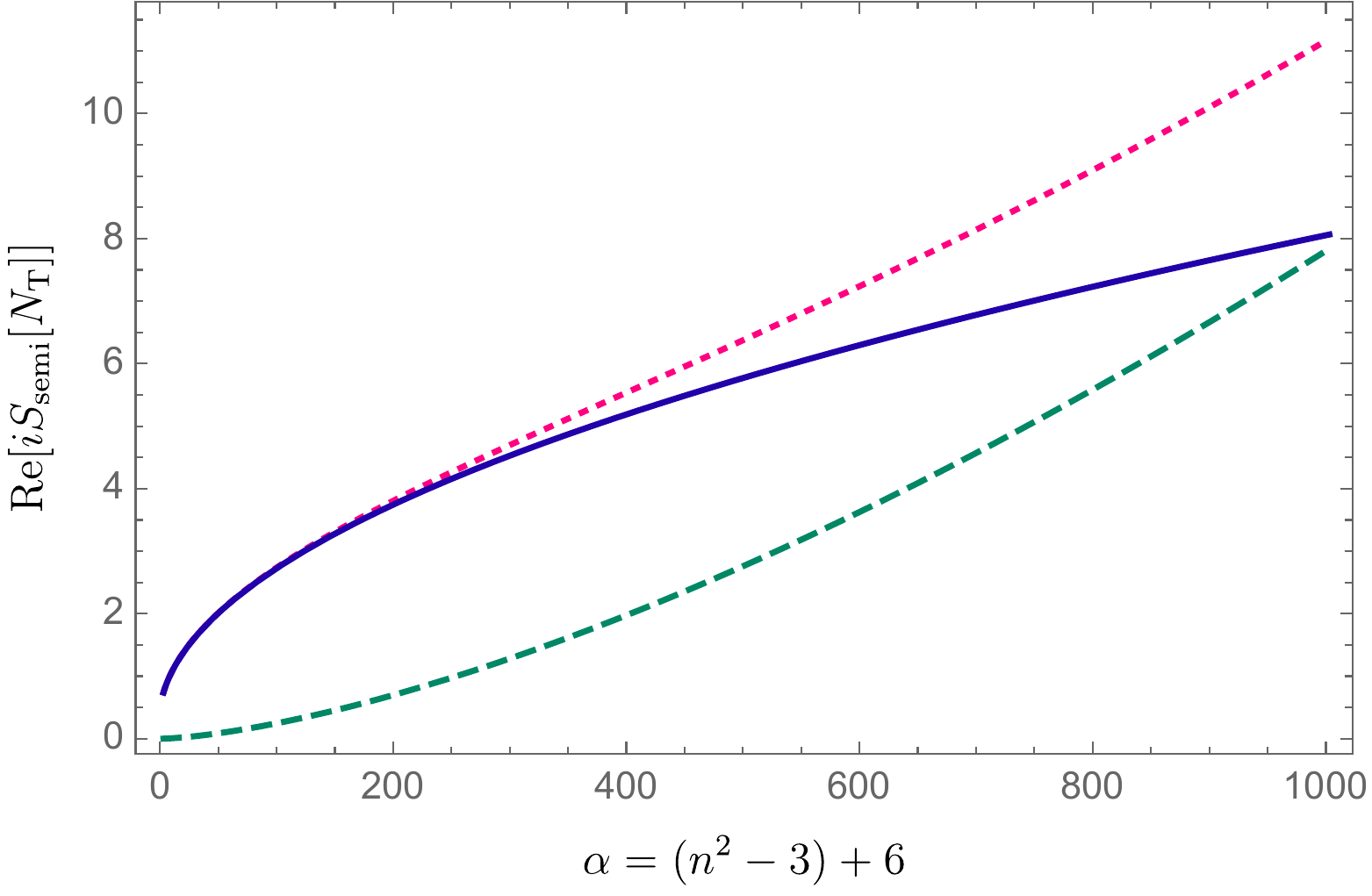}}%
          \subfigure[$H=10^{-3}$, $q_1=10^2$, ${\cal M}_{\rm UV}=1$]{%
		\includegraphics[clip, width=0.48\columnwidth]
		{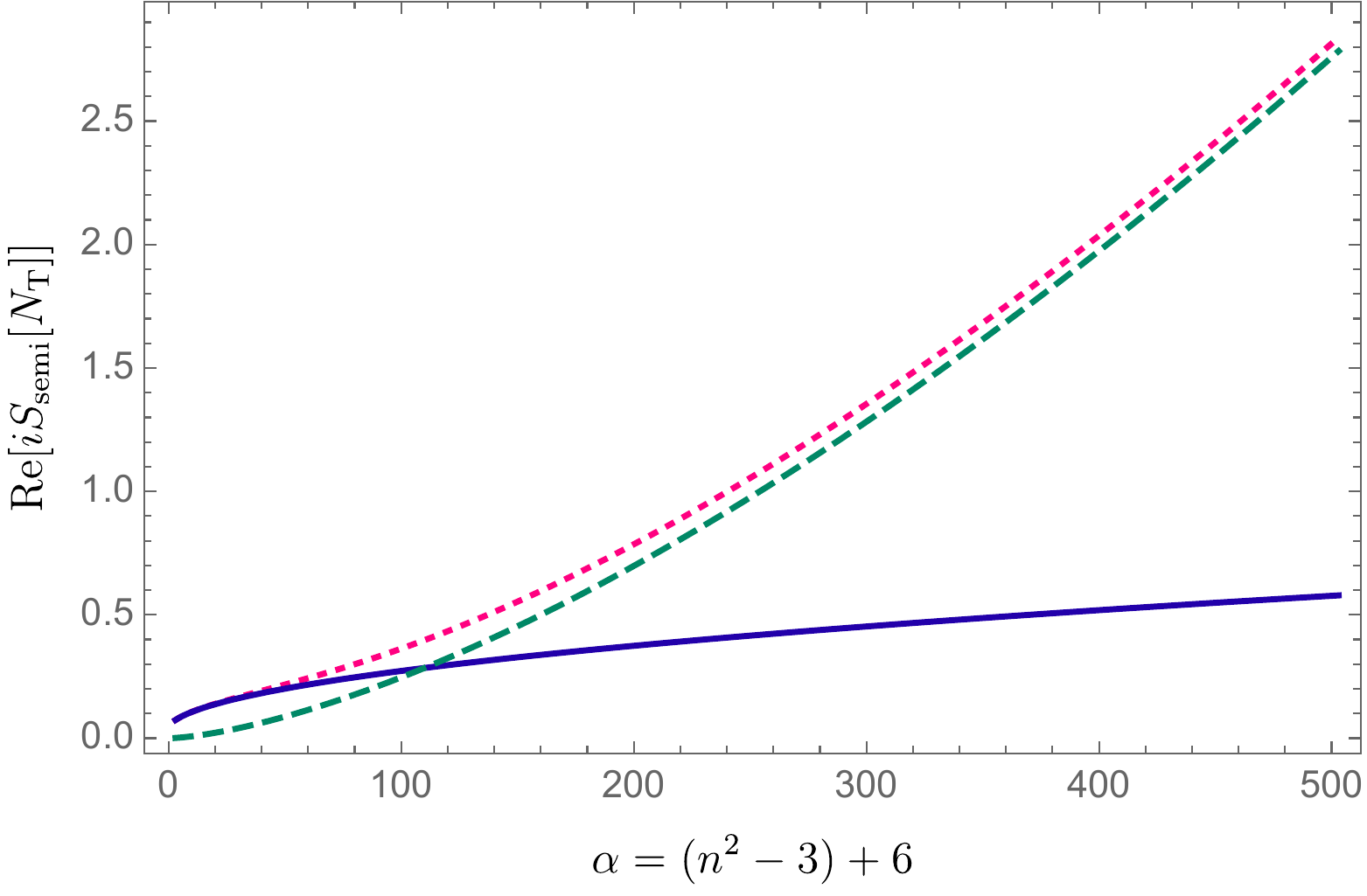}}%
	\caption{These figures show
$\textrm{Re}\left[iS_{\rm on-shell}[N_\textrm{T}]\right]$ at 
$N_\textrm{T}$~\eqref{eq:tunneling-saddle}
in terms of the mode $\alpha$ with $p=2$, $h_1=0.01$, $b_2=1$.
The blue solid and green dashed lines, respectively 
denote the GR formula~\eqref{eq:gr-semiclassical-action-ts}
and UV formula~\eqref{eq:gcj-semiclassical-action}  of
$\textrm{Re}\left[iS_{\rm on-shell}[N_\textrm{T}]\right]$ whereas 
the purple dotted line describes the numerical estimation of
$\textrm{Re}\left[iS_{\rm on-shell}[N_\textrm{T}]\right]$ given by solving 
the equation of motion~\eqref{eq:eom-jc} with the generalized Corley-Jacobson dispersion 
relation~\eqref{Modified-dispersion-general}. }
\label{fig:semiclassical-action-cj}
\end{figure}

From now on, we shall consider infrared (IR) modes with $\beta \ll q_1^2$ ($\alpha/q_1 \gg {\cal M}_{\rm UV}^2$ for $b_1=\mathcal{O}(1)$). In this case the behavior of $\chi$ near the UV ($\tau=0$) boundary is still given by (\ref{eq:gcj-solution}) and thus the regularity of the UV ($\tau=0$) contribution to the on-shell action requires $C_3=0$ for $\textrm{Re}[\lambda]<0$ and $C_4=0$ for $\textrm{Re}[\lambda]>0$. This gives the UV ($\tau=0$) boundary condition for (\ref{eq:eom-jc}). We solve the equation of motion~\eqref{eq:eom-jc} by imposing this boundary condition at the UV ($\tau=0$) boundary and another boundary condition $\chi(1)=h_1q_1$ at the IR ($\tau=1$) boundary. We then estimate $\textrm{Re}\left[iS^{(2)}_{\rm on-shell}[N_\textrm{T}]\right]$ at the tunneling saddle point~\eqref{eq:tunneling-saddle}.

The computation just outlined involves a numerical study, for which we first compute $\chi(\tau_i)$ and $\dot{\chi}(\tau_i)$ for sufficiently small $\tau_i$ ($0<\tau_i\ll 1$) using the UV formula (\ref{eq:gcj-solution}) with either $C_3=0$ (for $\textrm{Re}[\lambda]<0$) or $C_4=0$ (for $\textrm{Re}[\lambda]>0$), and use these values as the initial condition for the numerical integration of (\ref{eq:eom-jc}) towards larger values of $\tau$. We then numerically obtain $\chi(1)$ and $\dot{\chi}(1)$ up to a common overall factor, either $C_4$ or $C_3$. Finally, the on-shell action is numerically given by 
\begin{align} \label{eqn:onshellaction-numerical-howto}
 \frac{\pi^2}{4}q_1^2h_1^2 \frac{1}{N}\left(\frac{\dot{\chi}(1)}{\chi(1)} - \frac{N^2H^2+q_1}{q_1}\right),  
\end{align}
where the ratio $\dot{\chi}(1)/\chi(1)$ is independent of the overall factor ($C_4$ or $C_3$).

In Fig.~\ref{fig:semiclassical-action-cj}, we compare the numerical results of
$\textrm{Re}\left[iS^{(2)}_{\rm on-shell}[N_\textrm{T}]\right]$ with 
the analytical expression~\eqref{eq:gr-semiclassical-action-ts} in GR.
We set $\tau_i=10^{-5}$.
Even if the dispersion relation is modified by the generalized Corley-Jacobson dispersion relation in the UV region, we confirmed $\textrm{Re}[iS_{\rm on-shell}^{(2)}[N_\textrm{T}]]>0$ in the IR regime ($\alpha/q_1 \ll {\cal M}_{\rm UV}^2$),
and $S^{(2)}_{\rm on-shell}[N]$ is almost the same as the GR case. 
This is because for IR modes, the transition from the UV solution \eqref{eq:gcj-solution} to the IR solution \eqref{eq:gr-solution} takes place very close to the UV boundary $\tau=0$ and thus there is enough interval of $\tau$ for the growing IR solution to dominate the decaying IR solution before reaching the IR boundary $\tau=1$. Therefore, the ratio $\dot{\chi}(1)/\chi(1)$ in the expression \eqref{eqn:onshellaction-numerical-howto} for IR modes is essentially determined by the growing IR solution and thus insensitive to the boundary condition at the UV boundary unless the coefficient of the growing IR solution accidentally vanishes exactly. As a result, even for IR modes, the generalized Corley-Jacobson dispersion relation also shows the inverse Gaussian wave function and thus divergent correlation functions.

As a consistency check, Fig.~\ref{fig:semiclassical-action-cj} also shows numerical results for UV modes that agree with the analytical estimate \eqref{eqn:p=2inverseGorG} with $\textrm{Re}[\lambda]<0$.

\subsection{Unruh dispersion relation and trans-Planckian cutoff}

\begin{figure}
	\subfigure[$H=10^{-1}$, $q_1=10^2$, ${\cal M}_{\rm UV}=1$]{%
		\includegraphics[clip, width=0.48\columnwidth]
		{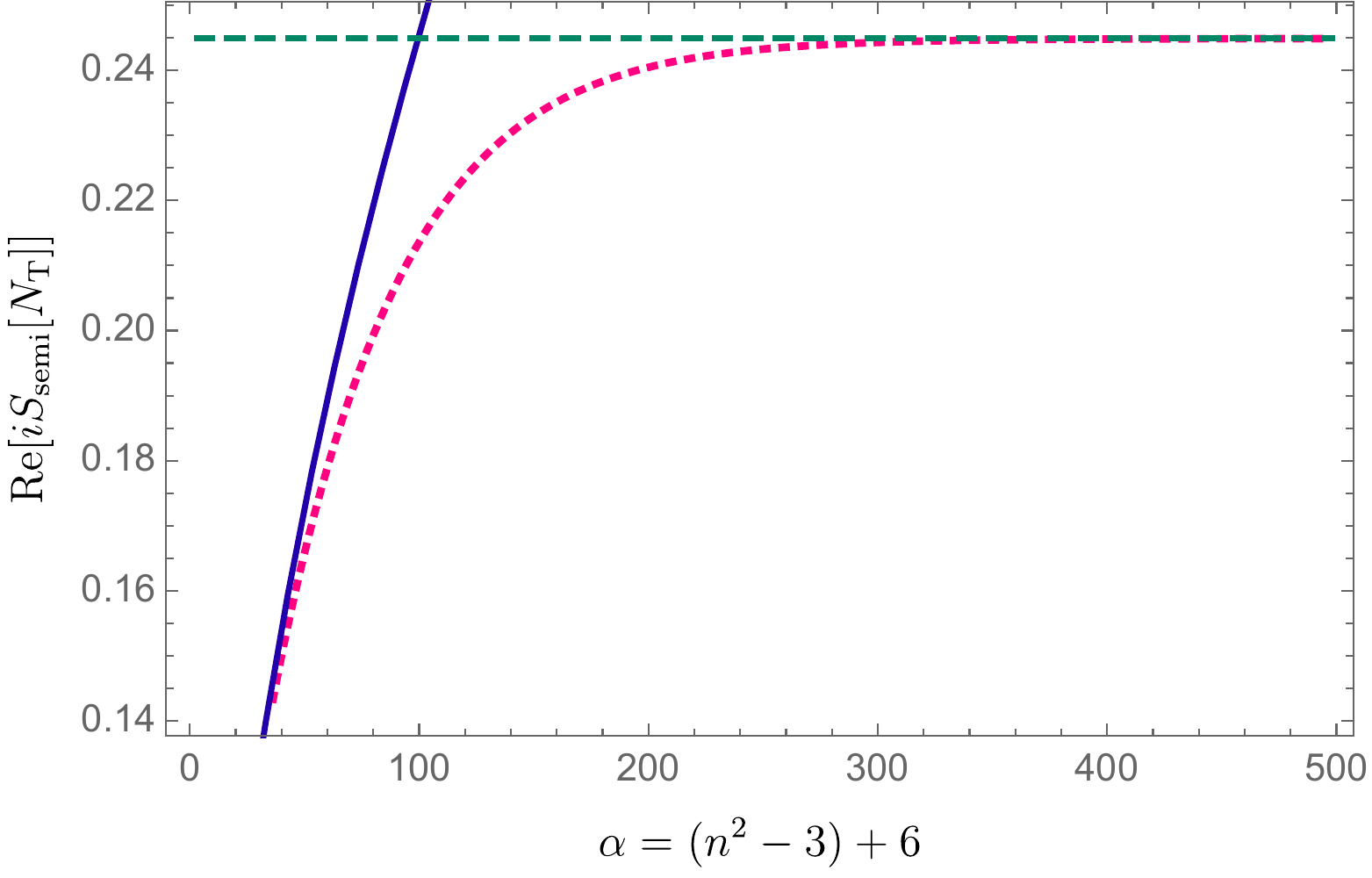}}%
	\subfigure[$H=10^{-1}$, $q_1=10^3$, ${\cal M}_{\rm UV}=1$]{%
		\includegraphics[clip, width=0.48\columnwidth]
		{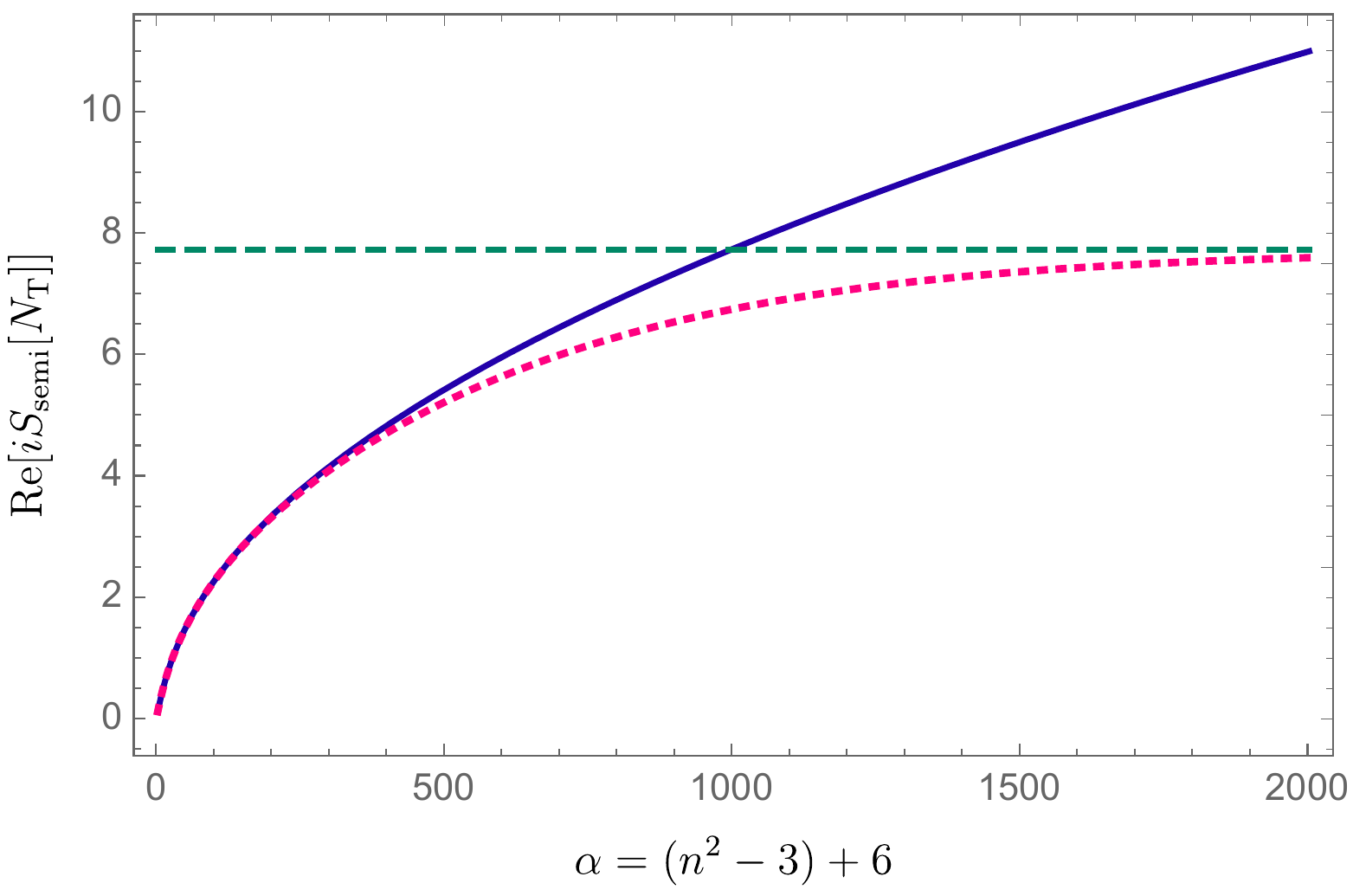}}
		\caption{
These figures show $\textrm{Re}\left[iS_{\rm on-shell}[N_\textrm{T}]\right]$ at 
$N_\textrm{T}$~\eqref{eq:tunneling-saddle}
in terms of the mode $\alpha$ with $h_1=0.01$.
The blue solid and green dashed lines denote 
the GR formula~\eqref{eq:gr-semiclassical-action-ts}
and UV formula~\eqref{eq:cutoff-semiclassical-action}  of
$\textrm{Re}\left[iS_{\rm on-shell}[N_\textrm{T}]\right]$ whereas 
the purple dotted line describes the numerical estimation of
$\textrm{Re}\left[iS_{\rm on-shell}[N_\textrm{T}]\right]$ given by solving 
the equation of motion~\eqref{eq:eom-jc} with the 
Unruh dispersion relation~\eqref{Modified-dispersion-Unruh}. }
\label{fig:semiclassical-action-Unruh}
\end{figure}

Next, we will consider 
the Unruh dispersion relation~\eqref{Modified-dispersion-Unruh} 
and discuss the regularity of the on-shell action for tensor perturbation 
with this dispersion relation.
First, to study the UV ($\tau=0$) contribution to the on-shell action, 
as an approximation that is valid near the UV boundary $\tau=0$ 
we shall utilize the trans-Planckian cutoff~\eqref{Modified-dispersion-cutoff}
instead of Unruh dispersion relation~\eqref{Modified-dispersion-Unruh}
and consider the following equation of motion,
\begin{align}
\frac{\ddot{\chi}}{N^2}  + \left\{ \frac{{\cal M}_{\rm UV}^2}{
H^2 N^2 (\tau -1) \tau +q_1 \tau}-\frac{2 H^2}{H^2 N^2 (\tau -1) \tau +q_1 \tau }\right\}\chi= 0 \,.
\end{align}
We obtain the following general solution,
\begin{align}\label{eq:cutoff-solution}
\chi(\tau)&=C_5\, G_{2,2}^{2,0}\left(
\begin{array}{c}
\frac{3-\Delta/H}{2},\frac{3+\Delta/H}{2} \\
 0,1 \\
\end{array}
|\frac{H^2 N^2 \tau }{H^2 N^2-q_1} \right) \notag\\
& \qquad -C_6\, \frac{H^2 N^2 \tau}{H^2 N^2-q_1}  
\, _2F_1\left(\frac{1-\Delta/H}{2},\frac{1+\Delta/H}{2};2\, ;\frac{H^2 N^2 \tau }{H^2 N^2-q_1}\right)
\end{align}
where $\Delta=\sqrt{9 H^2-4 {\cal M}_{\rm UV}^2 }$, Meijer G-function is defined by 
\begin{align}
G_{p, q}^{\, m, n}\left(\begin{array}{c}
a_{1}, \ldots, a_{p} \\
b_{1}, \ldots, b_{q}
\end{array} \mid z\right)=\frac{1}{2 \pi i} \int_{L} \frac{\prod_{j=1}^{m} \Gamma\left(b_{j}-s\right) \prod_{j=1}^{n} \Gamma\left(1-a_{j}+s\right)}{\prod_{j=m+1}^{q} \Gamma\left(1-b_{j}+s\right) \prod_{j=n+1}^{p} \Gamma\left(a_{j}-s\right)} z^{s} d s,
\end{align}
$_2F_1\left(a,b;c;z\right)$ is hypergeometric function,  
\begin{align}
F(a, b ; c ; z):={ }_{2} F_{1}\left[\begin{array}{c}
a, b \\
c
\end{array} ; z\right]=\sum_{n=0}^{\infty} \frac{(a)_{n}(b)_{n}}{(c)_{n}} \frac{z^{n}}{n !},
\end{align}
and the Pochhammer symbol is given by 
\begin{align}
(x)_{0}:=1, \quad (x)_{n}:=\prod_{k=0}^{n-1}(x+k).
\end{align}

By imposing the regularity of the UV ($\tau=0$) contribution to the on-shell action we have $C_5=0$.
Normalizing the overall factor of the solution~\eqref{eq:cutoff-solution} 
with $C_5=0$ by $\chi(1)=q_1h_1$, we obtain the following on-shell action for UV modes ($\alpha/q_1\gg {\cal M}_{\rm UV}^2$)
\begin{align}\label{eq:cutoff-semiclassical-action}
&S_{\rm on-shell}^{(2)}[N]
=\frac{\pi^2 Nh_1^2q_1}{8} \Biggl\{\frac{q_1 \left({\cal M}_{\rm UV}^2 -2 H^2\right)}{\left(H^2 N^2-q_1\right)} \frac{_2F_1\left(\frac{3-\Delta}{2},\frac{3+\Delta}{2};3;\frac{H^2 N^2}{H^2 N^2-q_1}\right)}{_2F_1\left(\frac{1-\Delta}{2},\frac{1+\Delta}{2};2;\frac{H^2 N^2}{H^2 N^2-q_1}\right)} -2 H^2 \Biggr\}
\end{align}
and confirm the inverse Gaussian distribution of the tensor perturbations at the tunneling saddle point $N_\textrm{T}$ at the UV regime $\alpha/q \gg {\cal M}_{\rm UV}^2$.

For IR modes ($\alpha/q_1\ll {\cal M}_{\rm UV}^2$), we numerically solve the equation of motion~\eqref{eq:eom2} with the Unruh dispersion relation~\eqref{Modified-dispersion-Unruh} in the same way as the previous case, 
and estimate $\textrm{Re}\left[iS^{(2)}_{\rm on-shell}[N_\textrm{T}]\right]$.
In Fig.~\ref{fig:semiclassical-action-Unruh}, we compare the numerical results for 
$\textrm{Re}\left[iS^{(2)}_{\rm on-shell}[N_\textrm{T}]\right]$ with 
the corresponding analytical GR expression~\eqref{eq:gr-semiclassical-action-ts}, 
and confirm that $S^{(2)}_{\rm on-shell}[N]$ is almost the same as the GR case for IR modes ($\alpha/q_1 \ll {\cal M}_{\rm UV}^2$). 
As a result, even in the IR region, the Unruh dispersion relation, including the trans-Planckian cutoff, does not solve the inverse Gaussian problem of the tensor perturbation.
As a consistency check, Fig.~\ref{fig:semiclassical-action-Unruh} also shows agreement between the numerical results and the analytical estimate \eqref{eq:cutoff-semiclassical-action} for UV modes.

\section{Conclusion}
\label{sec:conclusion}

We have investigated the problem of the inverse Gaussian wave function for perturbations in Lorentzian quantum cosmology, which describes the quantum creation of the universe from nothing. We have shown that this problem is inevitable as far as one requires the regularity of the contribution of the classical big-bang singularity to the on-shell gravitational action, and discussed whether the inverse Gaussian wave function for perturbations could be avoided by modifying the dispersion relation based on trans-Planckian physics. We have considered the generalized Corley-Jacobson dispersion relation and the Unruh dispersion relation as examples of the modified dispersion relation. The former dispersion relation can result from higher-dimensional operators in the gravity action. In the generalized Corley-Jacobson  dispersion relation, for $p=2$, we have found the analytical solution for the tensor perturbations for UV modes with $\alpha/q \gg {\cal M}_{\rm UV}^2$ and estimated the on-shell quadratic action for a tensor perturbation at the tunneling saddle point~\eqref{eq:tunneling-saddle}. We have found that the wave function leads to the inverse Gaussian distribution of the tensor perturbations, and UV modes of perturbations are out of control. We have numerically confirmed similar behavior for $p\neq2$. Also, for IR modes ($\alpha/q_1 \ll {\cal M}_{\rm UV}^2$), we have shown that the generalized Corley-Jacobson  dispersion relation does not solve the inverse Gaussian problem of the tensor perturbation. Indeed, for IR modes $S^{(2)}_{\rm on-shell}[N]$ shows excellent agreement with the GR case. For the Unruh dispersion relation, including the trans-Planckian cutoff, we have shown the same conclusion. Therefore, it is hard to overcome the problem of the inverse Gaussian wave function for tensor perturbations with the trans-Planckian physics modifying the dispersion relation in Lorentzian quantum cosmology.

\section*{Acknowledgment}
H.M. would like to thank Kazuhiro Yamamoto for his constructive comments. The work of H.M.~was supported by JSPS KAKENHI Grant No.~JP22J01284. The work of S.M.~was supported in part by JSPS Grants-in-Aid for Scientific Research No.~17H02890, No.~17H06359, and by World Premier International Research Center Initiative, MEXT, Japan. 
The work of A.N. was supported in part by JSPS KAKENHI Grant Numbers 19H01891 and 20H05852.

\bibliographystyle{JHEP}
\bibliography{reference}

\end{document}